\documentclass[prl,preprintnumbers,nofootinbib,twocolumn]{revtex4-1}
\pdfoutput=1
\usepackage{mathptmx}
\usepackage[english]{babel}
\usepackage{amssymb}
\usepackage{amsfonts}
\usepackage{amsmath}
\usepackage{eucal}
\usepackage{graphicx,color}
\usepackage{epsfig}
\usepackage{tikz}
\usepackage{cancel}
\usepackage{hyperref}

\usetikzlibrary{ arrows, babel, calc,decorations.pathmorphing, decorations.markings, decorations.pathreplacing, fit, positioning, shapes }

\newcommand{\be}{\begin{equation}} \newcommand{\ee}{\end{equation}}
\newcommand{\bea}{\begin{eqnarray}} \newcommand{\eea}{\end{eqnarray}}
\newcommand{\beann}{\begin{eqnarray*}}  \newcommand{\eeann}{\end{eqnarray*}}
\newcommand{\bfig}{\begin{figure}} \newcommand{\efig}{\end{figure}}
\newcommand{\ba}{\begin{array}} \newcommand{\ea}{\end{array}}
\newcommand{\bcen}{\begin{center}} \newcommand{\ecen}{\end{center}}
\newcommand{\btab}{\begin{tabular}} \newcommand{\etab}{\end{tabular}}

\usepackage{amsmath,mathtools,dsfont}
\newcommand{\id}{\mathds 1}
\DeclareMathOperator{\tTr}{tTr}

\def\barray{\begin{eqnarray}}
\def\earray{\end{eqnarray}}
\def\beq{\begin{equation}}
\def\eeq{\end{equation}}

\newtheorem{Proposition}{Proposition}[section]

\newtheorem{Theorem}{Theorem}[section]
\newtheorem{Lemma}{Lemma}[section]
\newtheorem{Corrolary}{Corrolary}[section]

\newcommand{\bp}{\begin{Proposition}}	\newcommand{\ep}{\end{Proposition}}
\newcommand{\bt}{\begin{Theorem}}	\newcommand{\et}{\end{Theorem}}
\newcommand{\bl}{\begin{Lemma}}		\newcommand{\el}{\end{Lemma}}
\newcommand{\bc}{\begin{Corrolary}}	\newcommand{\ec}{\end{Corrolary}}

\begin{document}

\title{Tensor renormalization group in  bosonic field theory}

\author{Manuel Campos, German Sierra and Esperanza Lopez}
\affiliation{
Instituto de F\'{\i}sica Te\'orica UAM/CSIC, C/ Nicol\'as Cabrera 13-15, Cantoblanco, 28049 Madrid, Spain}

\begin{abstract}
We  compute the partition function of a massive  free boson in a square lattice using a tensor network algorithm.
We introduce a  singular value decomposition (SVD) of  continuous matrices  that leads to
very accurate numerical results. It is shown the emergence of a CDL fixed point structure.
In the massless limit, we reproduce the results of conformal field theory including a precise
value of the central charge.
\end{abstract}

\preprint{IFT-UAM/CSIC-19-11}
\maketitle


Tensor Networks  (TN) have become in recent years a standard
technique    to study a wide variety  of problems in Condensed Matter Physics, Statistical Mechanics, Quantum Field Theory
 and other areas of Physics \cite{V08,O14}.
In quantum lattice systems TN  provide variational ansatzs for
many body wave functions denoted tensor network states (TNS).
Well known  examples of TNS are  Matrix Product States (MPS) for 1D systems \cite{A88,F92,K93,O95,V03,V04}
that underlies the DMRG method \cite{W92,D97,S05}, Projected Entangled Pairs States
(PEPS) that is a 2D version of MPS \cite{VC04,S98}, Multiscale Entanglement Renormalization Ansatz
(MERA) \cite{V07,G08,P09}, etc.
The use of TNS has also made possible
to classify the symmetry protected phases in 1D,  explore the  topological phases of matter in 2D  \cite{P10,C11,N11}
and provide simple versions of holography in the AdS/CFT correspondence
\cite{Sw12,LS15,PY15,Mo15,MN15,CK17}.

In classical spin systems the DMRG techniques where applied to compute the partition
function  \cite{N95}. Later on the method was improved expressing the partition function
and correlations using 4-index tensors \cite{M05}.
An important step was made by  Levin and Nave who  proposed the Tensor Renormalization Group (TRG) \cite{LN07}
were the  Kadanoff-Wilson blocking method is improved by implementing
entanglement techniques  in the truncation procedure \cite{K66,W75}.
However the TRG does not fully succeed in removing the short range entanglement.
For non critical systems, the TRG converges towards non trivial  tensors with  a {\em corner double line} (CDL) structure \cite{LN07,GW09}.
This difficulty was solved  by implementing techniques first developed for MERA \cite{EV15,EV15b}.

The aim of this letter is  to explore the application of real space tensor network techniques  to study quantum field theories.
Our motivation is to  revisit quantum field theory, and in particular renormalization group issues
from a framework naturally adapted to capture the role played by entanglement. As a first step, we efficiently adapt the
TRG protocol to evaluate the partition function of a free boson.
Like in the ordinary TRG, a CDL type infrared fixed point emerges at the expected length scale. In the conformal limit we obtain a competitive
estimation for the value of the central charge.
Our implementation of the TRG is based on the simple rules of gaussian integration, and hence
we name it gaussian TRG (gTRG).

{\it The model}.
We will consider a free scalar of mass $m$ in two dimensions.
Continuous versions of tensor networks have been proposed for the study of quantum field theories  \cite{V10,H13,J15,TC18,HV18}.
However they are not yet developed to the extent ordinary tensor networks are, and we will not pursue them here.
In the following,  space-time will be discretized while field variables retain their continuous character.
This choice breaks  symmetries like translation and rotation but they can be recovered in the continuum limit.
Space-time will be represented by a square lattice with periodic boundary conditions.
At each site $(i,j)$  of the lattice lives a variable $\phi_{ij} \in \mathbb{R}$. The euclidean partition function  is
\be
Z= \int \prod_{ij} d \phi_{ij} \; e^{-{1 \over 2} \; \sum_{ij} \left[  (\phi_{ij}-\phi_{i+1 j})^2 \,+\,  (\phi_{ij}-\phi_{i j+1})^2 \,+\, m^2 \phi_{ij}^2 \right]} \; ,
\label{lZ}
\ee
where $m$ is measured in lattice units.

The interactions on  the lattice described by \eqref{lZ} are pairwise between the fields at neighbour sites.
It is convenient to change to a vertex model, where the fields live on the edges and
the interactions take place at the lattice sites.
 On the dual tilted lattice, we define the statistical weights
\be
W({\phi_i})= e^{-{1 \over 2} \sum_{i=1}^4 \big[(\phi_i -\phi_{i+1})^2 \, + \, {m^2 \over 2} \phi_i^2\big] } \, .
\label{W}
\ee
that can be depicted as
\begin{equation}
	  \vcenter{\hbox{\includegraphics{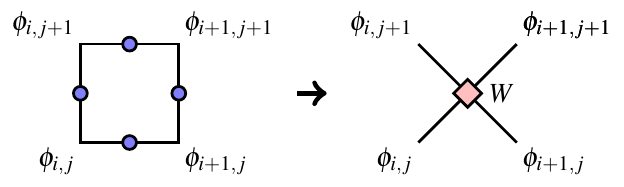}}}
\nonumber
\end{equation}
\noindent
We have shadowed the interaction vertices for clarity.

{\it Gaussian SVD}.
We will implement a TRG protocol to reduce iteratively  the number of degrees of freedom.
The basic tool used in systems with a finite number of degrees of  freedom is the
singular value decomposition (SVD) of the network tensors. Any finite rank matrix can be decomposed as $M=U S V^\dag$,
where $U$ and $V$ are unitary matrices and $S$ is diagonal with non-negative entries.
The latter result
also holds for  compact operators acting on  Hilbert spaces  of continuous functions.
This result has been used  to implement the standard TRG approach to a  $\phi^4$-boson field theory  \cite{Sh12}.
Here we shall not follow this approach but one that is inspired  on standard field theory techniques.
Indeed, we will impose  two  requirements at each step of the coarse graining procedure: i)  the statistical weights should remain gaussian
and ii) the lattice variables should be  continuous fields. These requirements leads us to adapt  the  SVD suitably.

We will allow several fields to live at each  lattice edge. For simplicity we still denote them collectively as ${ \phi}\equiv \{\phi_1,..,\phi_\chi\}$. The number of fields per edge plays the role of bond dimension. We group the fields entering each vertex in two sets  labelled  as $L$ and $R$.
Generic gaussian weights have the form
\vspace{-2mm}
\be
\vcenter{\hbox{\includegraphics{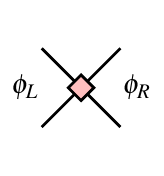}}}
\hspace{1mm}= \hspace{1mm}
\rho \, e^{- {1 \over 2}{ \phi}_L^{\;T}  A_L  { \phi}_L- {1 \over 2}{ \phi}_R^{\;T\,} A_{R \,} { \phi}_R +{ \phi}_L^{\;T}  B \, { \phi}_R} \; ,
\label{expW}
\ee
\vspace{-5mm}

\noindent with $A_{L,R}$ and $B$ real matrices of dimension $2\chi \times 2 \chi$ and $\rho$ a constant. We search for a decomposition of $W$ inspired in the SVD.
Namely, we want to factorize the dependence on L and R fields by introducing new variables, which according to the previous requirements should have the interpretation of fields
\begin{equation}
  \vcenter{\hbox{\includegraphics{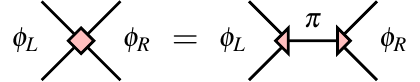}}}
\label{TRGfig}
\nonumber
\end{equation}

A way to proceed is working directly with the quadratic forms that appear in
the exponent of the gaussian weights.
The L and R fields are connected by the matrix B, which thus hinders factorization.
Since $B$ is real, we have $B=UD \,V^T$ with $U$ and $V$ also real.
We are assuming that $D$ contains only strictly positive entries and hence it is of dimension ${\tilde \chi}= \rm{rank} (B)\leq 2\chi$. Introducing $\tilde \chi$ new fields ${ \pi}$,
we can rewrite
\be
W({ \phi}_L,{ \phi}_R)= G_L ({ \phi}_L) \; {\widehat W} ({ \phi}_L,{ \phi}_R) \;  G_R ({ \phi}_R) \, ,
\label{gaussianSVD}
\ee
where we have used straightforward gaussian integrations to define
\barray
{\widehat W}  &  =  &  \int d { \pi} \; e^{\, i \,{ \phi}_L^{\;T}  U{ \pi}} \;  S({ \pi})\;  e^{- i \, { \pi}^{\;T}  V^T  { \phi}_R} \, ,
\label{svd1}
\\
S & = &  {1\over \sqrt{(2 \pi)^{\tilde \chi} \det \,D}} \; e^{-{1 \over 2}{ \pi}^{\;T} D^{-1} { \pi}} \; . \hspace{.9cm}
\nonumber
\earray

Relation \eqref{svd1} is a  continuous SVD,
with the entries of the diagonal matrix $S$ providing the singular values. ${ \pi}$ act as canonically conjugate variables of the
original fields. However, the  diagonal factors $G_{L,R}$ cause \eqref{gaussianSVD} to deviate from a SVD
\be
G_L= e^{-{1 \over 2}{ \phi}_L^{\;T}\, (A_L - U D  U^T) \,{ \phi}_L }  \;, \;\;\;\;  G_R= e^{-{1 \over 2}{ \phi}_R^{\;T}\, (A_R - V D  V^T) \,{ \phi}_R } \, .
\label{WLR}
\ee
These matrices will probe crucial in the implementation of the TRG.
They are the price to pay for the enormous simplification of working at the level of the exponent, dealing only
with finite dimensional matrices. We will refer to \eqref{gaussianSVD} as gaussian SVD (gSVD).

{\it Gaussian TRG}. It is an  iterative application of the following  transformations
of a model defined  on a lattice of  $N$ sites into a lattice of  $N/2$ sites
\vspace{-1mm}
\begin{equation}
  \vcenter{\hbox{\includegraphics{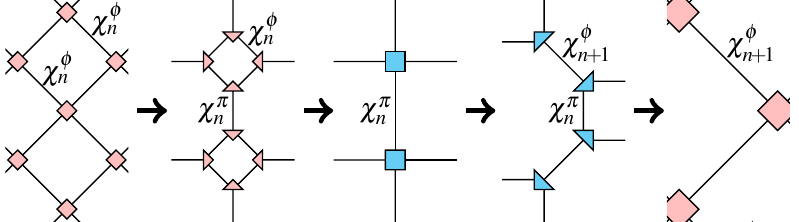}}}
\label{TRGfig}
\nonumber
\end{equation}
namely:
i) gSVD  of  the weights of  the $\phi$-fields,
ii) construction of the weights of the $\pi$-fields,
iii) gSVD  of  the weights of  the $\pi$-fields,
and iv) construction   of  the weights of the   $\phi$-fields.
The $\phi$ and $\pi$-fields turn out to have very different properties (see below). We shall label the associated matrices with a subscript $\phi$ or $\pi$,
corresponding to the tilted and directed lattices that are rotated by $45^\circ$ every TRG transformation.
A complete  RG cycle returns  to the same type of lattice, and thus it is  composed of two TRG steps.

We will use a subindex $n$ to label the RG iteration as indicated in the above figure.
The initial lattice, defined by the weights \eqref{W},  has by assumption
${\chi_1^{}}^{\!\!\!\phi}=1$. Its associated  matrix $B^\phi$ has two equal singular values, and thus $\chi_1^\pi=2$.
With no truncations,  the bond dimension doubles when transforming from $\phi$ to $\pi$-fields, i.e.
$\chi^\pi_n = 2 {\chi_n^{}}^{\!\!\!\phi}$, and remains constant in the reverse step, i.e. $\chi_n^\pi = \chi_{n+1}^\phi$.
Hence ${\chi_n^{}}^{\!\!\!\phi} = 2^{n-1}$.

The singular values added  at each RG transformation are expected to encode correlations at larger
coarse grained scales.
In the vacuum of the bosonic theory correlations decay with distance.
Hence at some RG step  the new singular values should start being sufficiently small to set them to zero with a small error cost.
This reduces the dimension of the ancillary  field space and renders the calculation feasible.
Since we are not dealing with an ordinary SVD, there is some degree of ambiguity involved in this
implementation. We will proceed as follows. The matrix $B$ can be rewritten as
\be
B=U_1 D_1 V_1^T + U_2  D_2 V_2^T \, ,
\ee
where $D_i$ are
diagonal matrices with the highest ($i=1$), and smallest ($i=2$), eigenvalues of $B$ respect to
a chosen cutoff.
Based on that, we can substitute
\be
S({ \varphi}) \rightarrow \, S_1 ({ \varphi}_1) \delta({ \varphi}_2)  \, ,
\label{Wapprox}
\ee
where ${ \varphi}=\{{ \varphi}_1,{ \varphi}_2\}$ and $\varphi$ can refer to the $\phi$ or $\pi$-fields.
The matrix $S_1$ is given by \eqref{svd1} with $D$ replaced by $D_1$.
The delta function eliminates the dependence on the fields ${ \varphi}_2$, reducing the bond dimension.
The difference between the exact and  the truncated weights is $\Delta W= G_L \, {\widehat W} \Delta {\widehat W} \, G_R$,
where
\be
\Delta {\widehat W} = 1-e^{\,{1 \over 2}({ \phi}_L^T U_2 - { \phi}_R^T V_2)\, D_2\, (U_2^T { \phi}_L-V_2^T { \phi}_R)} \, .
\ee
In the large field limit, $\Delta {\widehat W}$ can be arbitrarily large no matter how small are the entries of $D_2$.
In order to justify \eqref{Wapprox} it is necessary to have the large field values suppressed.
This is  achieved by   the factors $G_{L,R}$, which in particular contain the mass terms for
the lattice fields. The high accuracy of the numerical  results
presented  below indicates that these matrices indeed play efficiently the role of field regulators.

We name this adapted TRG protocol gTRG. The integration leading to the new weights at each gTRG step are gaussian and thus easy to perform. From now on we use a scheme in which $U=V$,
and hence
the relation $A_L=A_R$ satisfied by the initial weights will be preserved (see \hyperref[sec:SMA]{SM.A}). In this scheme the same gSVD data characterize every lattice site.

\begin{figure}[h]
\begin{center}
\includegraphics[width=5.8cm]{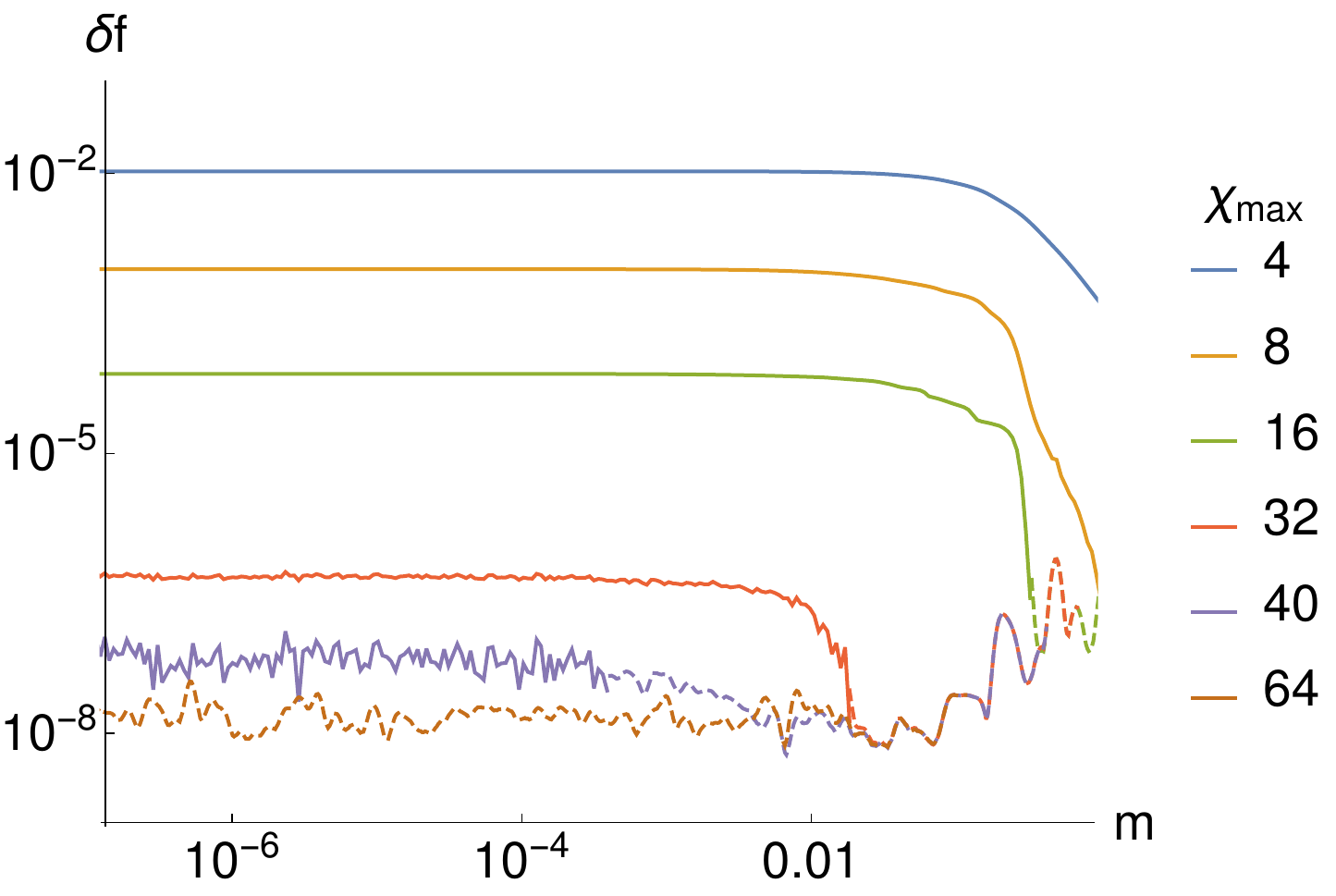}~~~
\end{center}
\vspace{-5mm}
\caption{\label{fig:fe} Relative error in the free energy per site $\delta f$  as a function of the mass $m$, for a
 lattice with $L_1=L_2=2^{30}$, and maximal bond dimensions $\chi_{\rm max}$.}
\end{figure}

{\it Results}. The partition function of a free boson can be computed analytically  using momentum eigenmodes.
For a lattice of size $L_1 \times L_2$  with periodic boundary conditions it reads
 \be
Z^{\rm exact}_{L_1L_2} =   \Big( \frac{\pi}{2} \Big)^{ \!\!\frac{L_1 L_2}{2} }
\! \prod_{n_1, n_2 } \!\!  \left( \sin^2 \frac{\pi n_1}{L_1} +  \sin^2 \frac{\pi n_2}{L_2}  +  \frac{m^2}{4} \right)^{\!\!- {1\over 2}}
 \label{zexact}
 \ee
 where $n_i=1, \dots, L_i \;(i=1, 2)$. Comparison with the exact result allow us to
test the performance of the  gTRG method.
In Fig.\ref{fig:fe} we plot the
relative error $\delta f$ in the free energy per site,  $f = -  \ln Z/L_1 L_2$,
as a function of the mass for different maximal bond dimensions $\chi_{\rm max}$. A large lattice with $L_1=L_2=2^{30}$
has been chosen.
With $\chi_{\rm max}=32$ we obtain an error below $10^{-6}$. The results for $\chi_{\rm max}>32$ become increasingly noisy because we reach the accuracy limit of the numerical tools we are using: Mathematica with default settings. The dashed lines in Fig.\ref{fig:fe} are averaged results for the absolute value of $\delta f$. With $\chi_{\rm max}=64$ the average precision is $10^{-8}$, while in the best cases we have reached an error below $10^{-9}$.

\begin{figure}[h]
\begin{center}
\includegraphics[width=4.2cm]{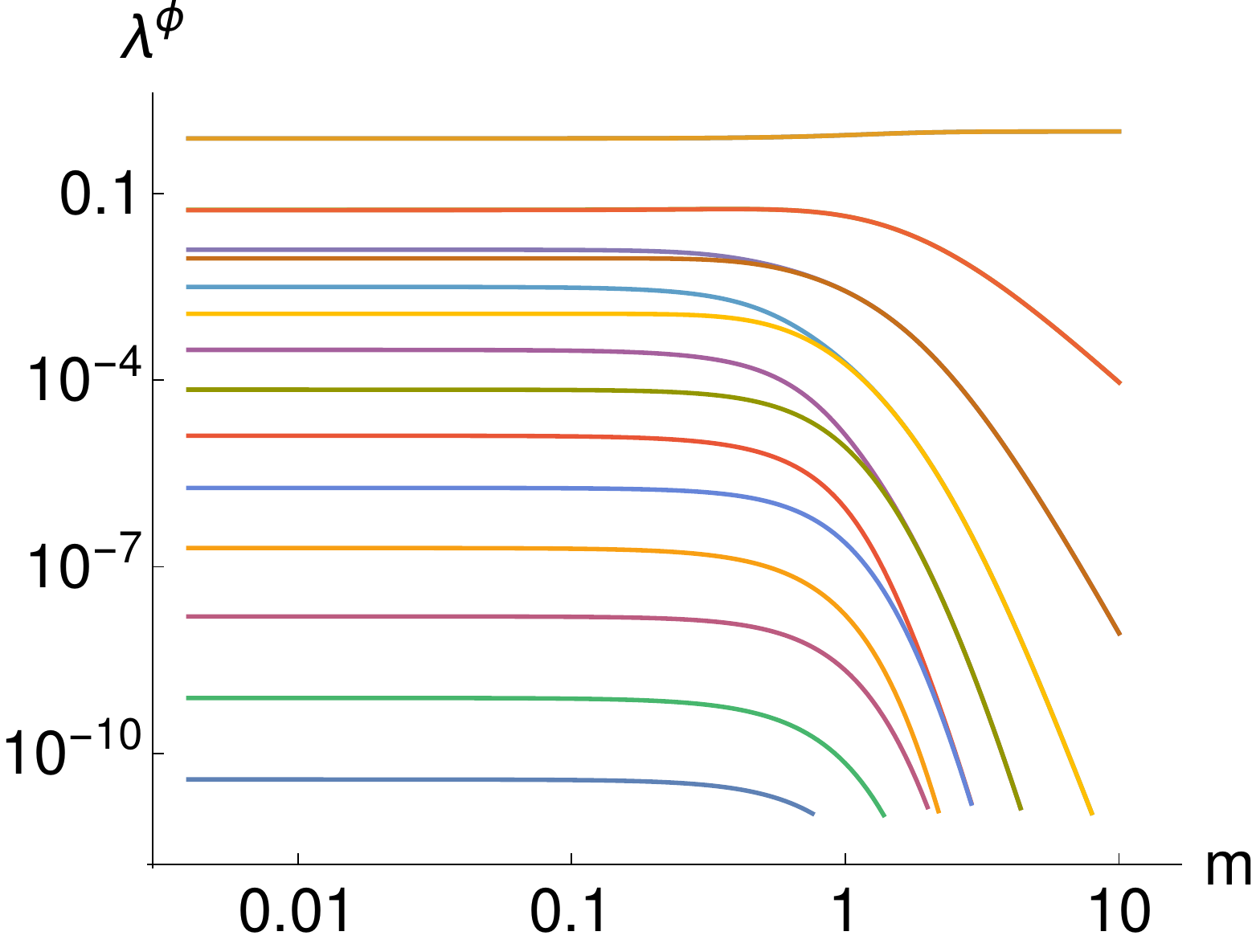}~~~
\includegraphics[width=4.2cm]{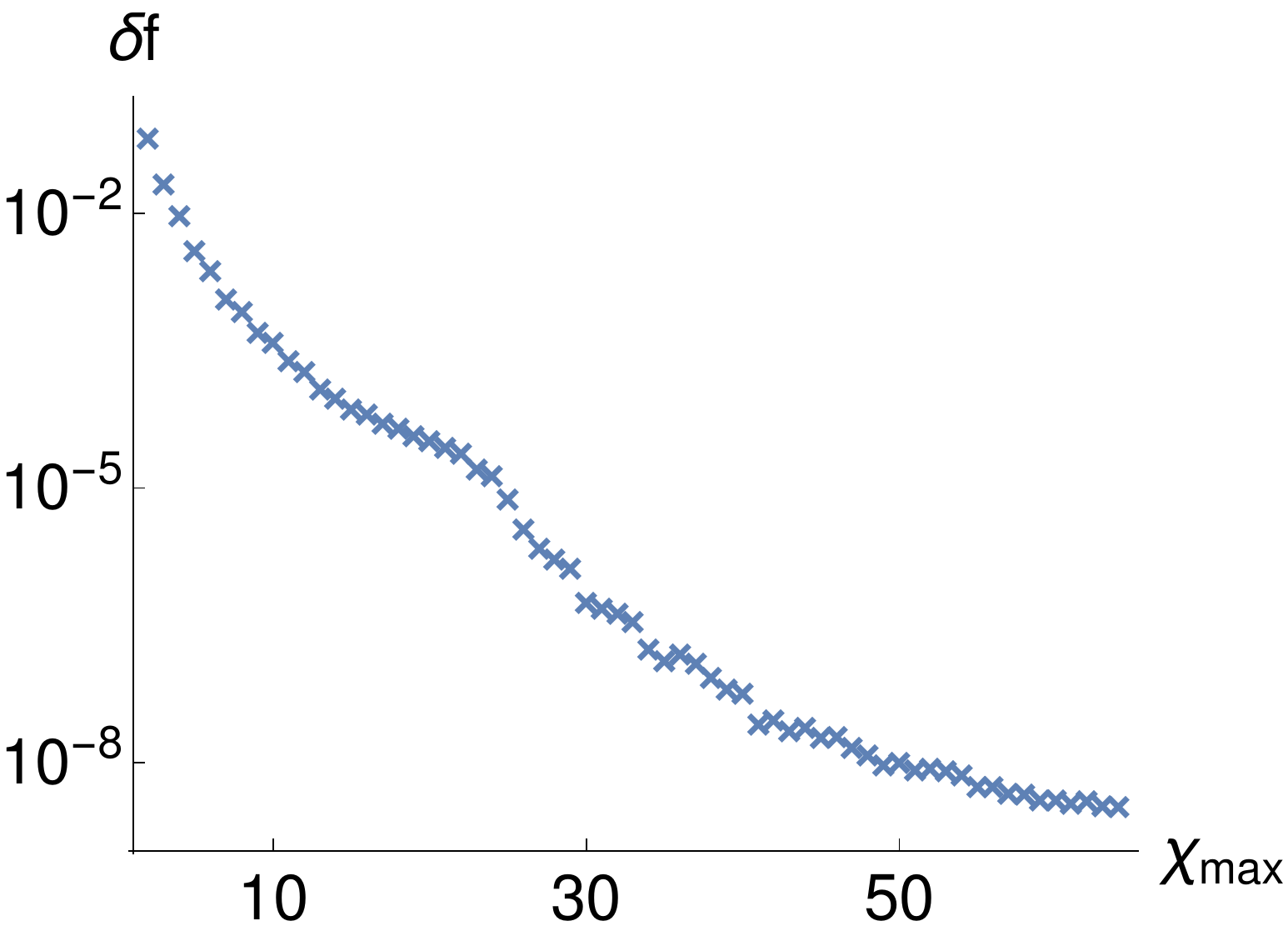}
\end{center}
\vspace{-5mm}
\caption{\label{fig:svB}
Left: Singular values of $B^\phi_4$ with no truncation.
For large masses they always join in equal value pairs.
The two top curves correspond to doubly degenerate singular values.
Right: $\delta f$ for $m= 1.2 \times 10^{-6}$ as a function of the bond dimension for $L_1=L_2=2^{30}$.}
\end{figure}

Truncation is introduced in a step leading from a $\phi$ to a $\pi$-lattice, since it is then when the bond dimension increases.
Fig.\ref{fig:svB}-left shows the singular values of $B^\phi_4$. No truncation has been yet applied and hence ${\chi_4^{}}^{\!\!\! \phi}=8$.
We observe that the singular values are very strongly decaying. This general property allow us to truncate them affecting only mildly
the accuracy  of the results. Notably it also holds in the limit of very small masses, explaining the smooth and efficient behaviour of the gTRG in a regime which is problematic for the ordinary TRG.

Independently of the bond dimension, we have discarded singular values smaller than a threshold $\epsilon$ in order to minimize numerical errors.
The value of $\epsilon$ depends on the numerical precision with which we are operating. In our case, we found appropriate to set $\epsilon=10^{-11}$.
Imposing this threshold in fact improves the effectiveness of the TRG in a rather not trivial way that involves both $\phi$ and $\pi$-fields and
is explained in the \hyperref[sec:SMB]{SM.B}.
Fig.\ref{fig:svB}-right shows the relative error in the free energy per site as a function of the bond dimension for $m=10^{-6}$. This curve has two well differentiated segments. The first one falls as
$\chi^{-a}$, with $a\sim 3.44$. This is the typical TRG behaviour, in which improving the precision is increasingly expensive \cite{LN07,EV15}.
The parameter
$\epsilon$ starts playing a role at $\chi_{\rm max}=22$.
At this point the curve enters its second segment, where we observe that the precision improves at a lower computational cost.

 {\it Massless case.-}  The accurate results of the gTRG for small masses allow us to address the massless case.
 In the limit $m\ll 1$ and $L_1,L_2 \gg 1$, with $L_2/L_1$ constant, the exact partition function \eqref{zexact}
 can be approximated by (see \hyperref[sec:SMD]{SM.D})
 \be
Z^{\rm \, exact} _{L_1 L_2}  \simeq     \frac{  e^{  -   f_\infty L_1 L_2  } }{ m ( L_1 L_2)^{ 1/2} }
\;\;  Z_{\rm CFT}(\tau)     \, ,
 \label{56}
 \ee
 where $Z_{\rm CFT}$ is the partition function of
 a massless boson in a torus with moduli parameter $\tau$  \cite{CFT}.
In our case $\tau =  i L_2/L_1$.

The leading contribution to the free energy per site comes from the exponential term in \eqref{56}
\be
f_\infty = \frac{ 2 G}{\pi}  - \frac{ \ln( 2  \pi)}{2} \, ,
\label{57}
\ee
where $G$ is the Catalan constant. The CFT partition function is responsible for the leading finite size corrections.
Choosing $L_1 = L_2 \equiv L$, equation (\ref{56}) yields
\be
\frac{\pi}{6} c_{\rm th} = L^2( f_\infty - f) +  \ln ( mL) + 2 \sum_{n=1}^\infty \ln ( 1 - e^{-2\pi n}) \, ,
\label{57b}
\ee
where $c_{\rm th}=1$ is the theoretical value of the central charge \cite{A86,B86}.
Taking $L= 2^6$ and  $2^{7}$ and using  (\ref{57b}) we obtain respectively
\be
c_{\rm gTRG} - c_{\rm th}= O(10^{-5}), \;O(10^{-6}) \, .
\ee
These values are derived with $\chi_{max}=64$ by averaging over $m\in [ 10^{-14},10^{-8} ]$ in order to minimize the numerical noise.
For larger $L$ the numerical noise wins over the leading finite size effect, while for smaller lattices higher order finite size effects worsen the result.

{\em RG flow.} The RG behaviour of free field theories is extremely simple.
When a mass parameter is present, it runs with the scale according to its bare dimension.
Hence a small mass will become of order one in lattice units after
\be
n(m) \sim -  \frac{ \log m}{\log 2}  \, ,
\label{cdl}
\ee
RG iterations.
For $n \gtrsim n(m)$ correlations should be mostly confined to occur inside a single lattice plaquette.
Entanglement inside a plaquette is modelled by a corner double line (CDL) structure \cite{LN07,GW09}
\begin{equation}
 \vcenter{\hbox{\includegraphics{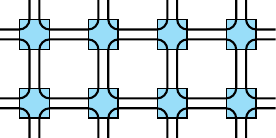}}}
	\nonumber
\end{equation}

\noindent
The TRG has the drawback of being unable to eliminate such ultralocal entanglement and reach a trivial IR fixed point.
Instead it promotes the inner correlations from half of the plaquettes to the next coarse graining level, reproducing again a CDL structure. The same should apply to the gTRG.

\begin{figure}[h]
\begin{center}
\includegraphics[width=4cm]{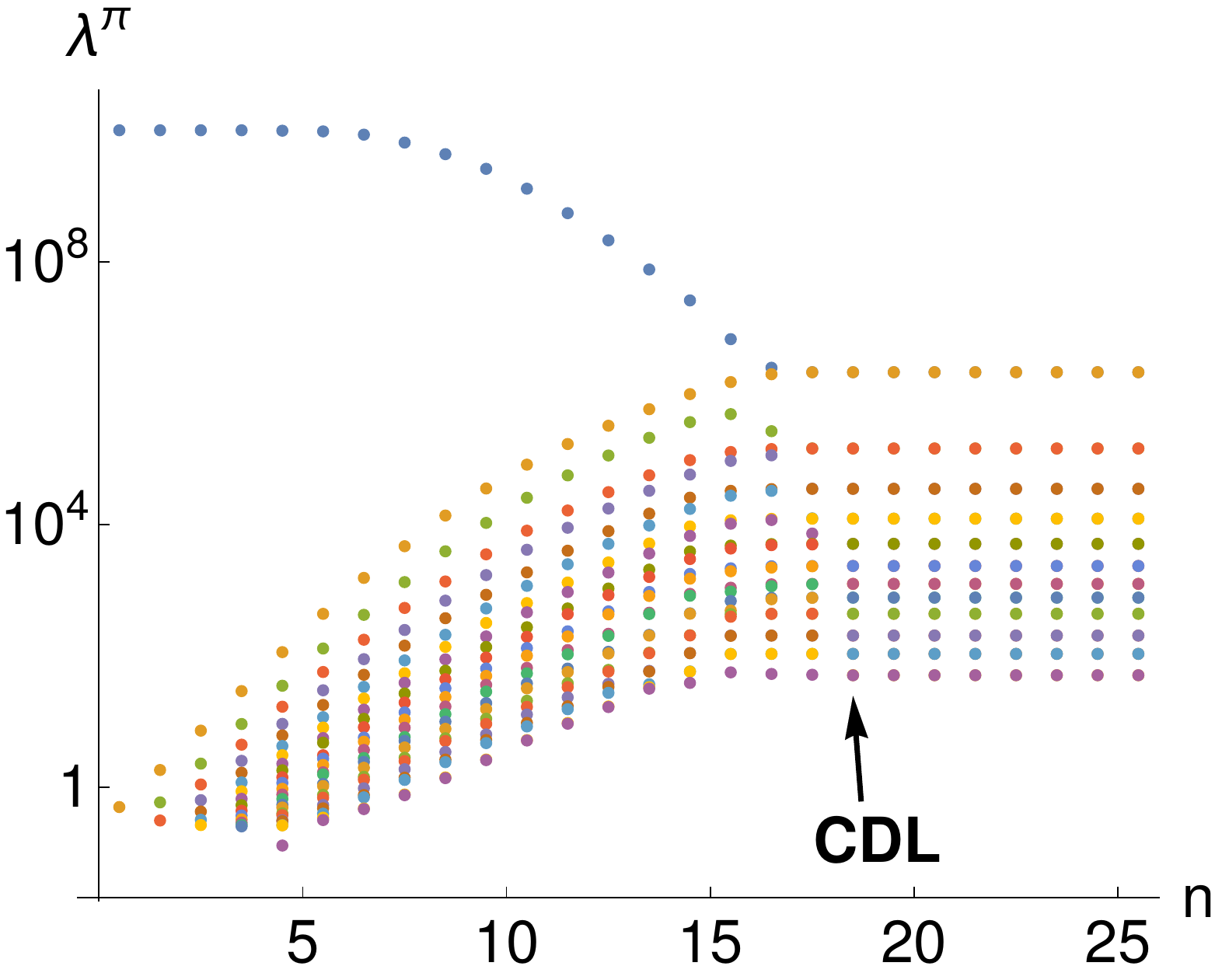}~~~
\includegraphics[width=4.cm]{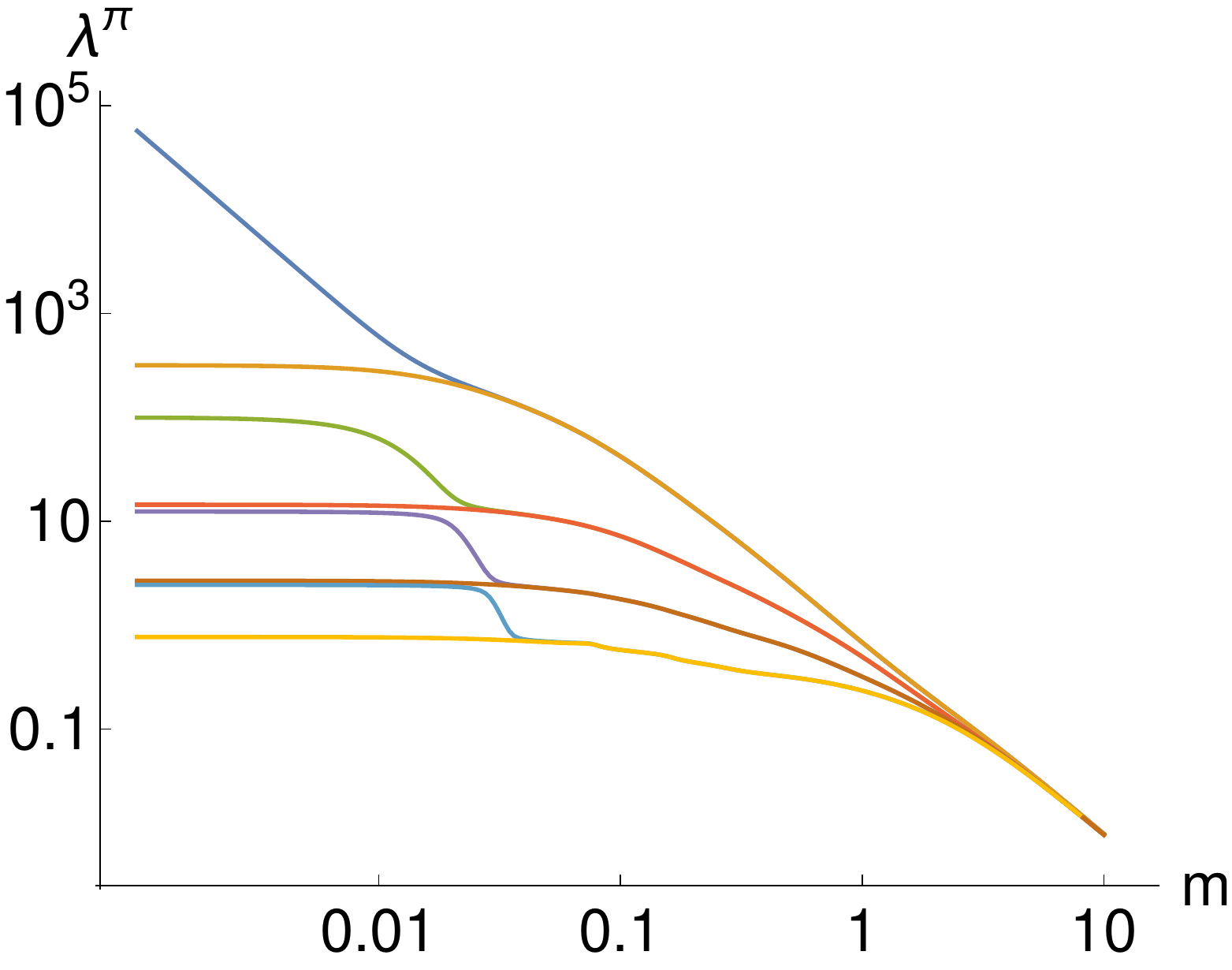}
\end{center}
\vspace{-2mm}
\caption{\label{fig:CDL1} Singular values of $B^\pi_n$. Left:  RG flow for $m=10^{-5}$ and $\chi_{\rm max}=24$.
Right: As a function of the mass, for $n=8$ and $\chi_{\rm max}=8$.}
\end{figure}

The emergence of a CDL structure requires that the singular values of $B$ form equal value pairs.
The singular values of $B^\phi$ have a strong tendency to arrange in pairs. Indeed, Fig.\ref{fig:svB}-left shows that the six highest
singular values have already paired up after three RG cycles.
This is however not the case for $B^\pi$.
Its singular values in the first RG cycle can be derived explicitly
\be
\lambda_1^\pi = {1 \over m^2} \,{8 + 4 m^2 + m^4 \over 8  + 6 m^2 + m^4}\; , \hspace{.5cm} \lambda_2^\pi={1 \over 2 + m^2} \; .
\ee
For small masses $\lambda_1^\pi \approx 1/m^2$ and $\lambda_2^\pi \approx 1/2$. In successive RG cycles,
the gap between the largest singular value and the rest slowly decreases until it closes.
The singular values then pair up as required for CDL behaviour and acquire fixed values. The smaller the mass, the larger the gap and the
more RG iterations are necessary. Fig.\ref{fig:CDL1}-left shows the RG flow of the singular values for $m=10^{-5}$ and $\chi_{\rm max}=24$.
Pairing is effective for $n \simeq 19$ in agreement with \eqref{cdl}, which gives $n(10^{-5}) \simeq 16-17$.

The same behaviour is seen in Fig.\ref{fig:CDL1}-right.
We have plotted the singular values of $B^\pi_8$ obtained with $\chi_{\rm max}=8$. The singular values pair up for masses larger than $m\approx 0.03$. Below they rapidly unpair, with the largest singular value strongly detaching from the rest. In rescaled lattice units the threshold mass is $0.03 \times  2^8 \approx 8$.
Hence a CDL structure does not emerge until scales larger than the correlation length, $\xi=1/m$, are reached.

\begin{figure}[h]
\begin{center}
\includegraphics[width=4.4cm]{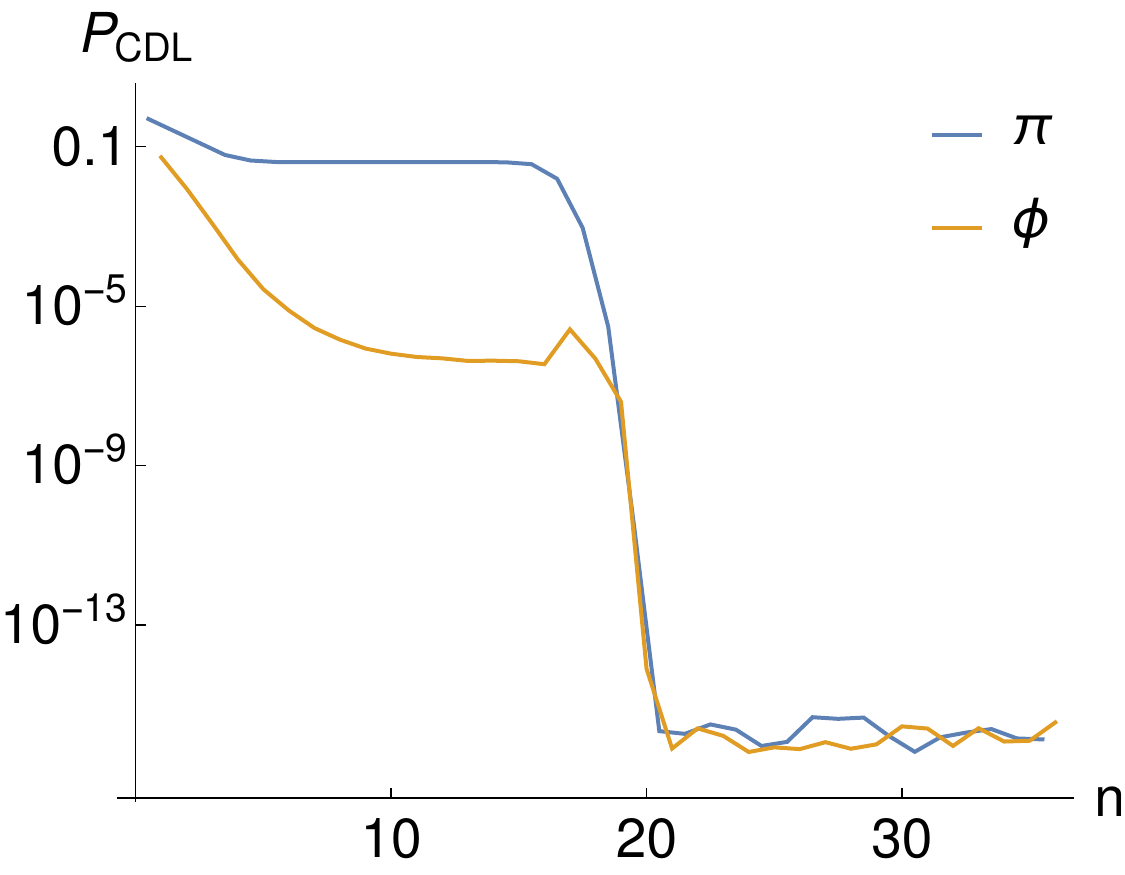}~~~
\includegraphics[width=4.0cm]{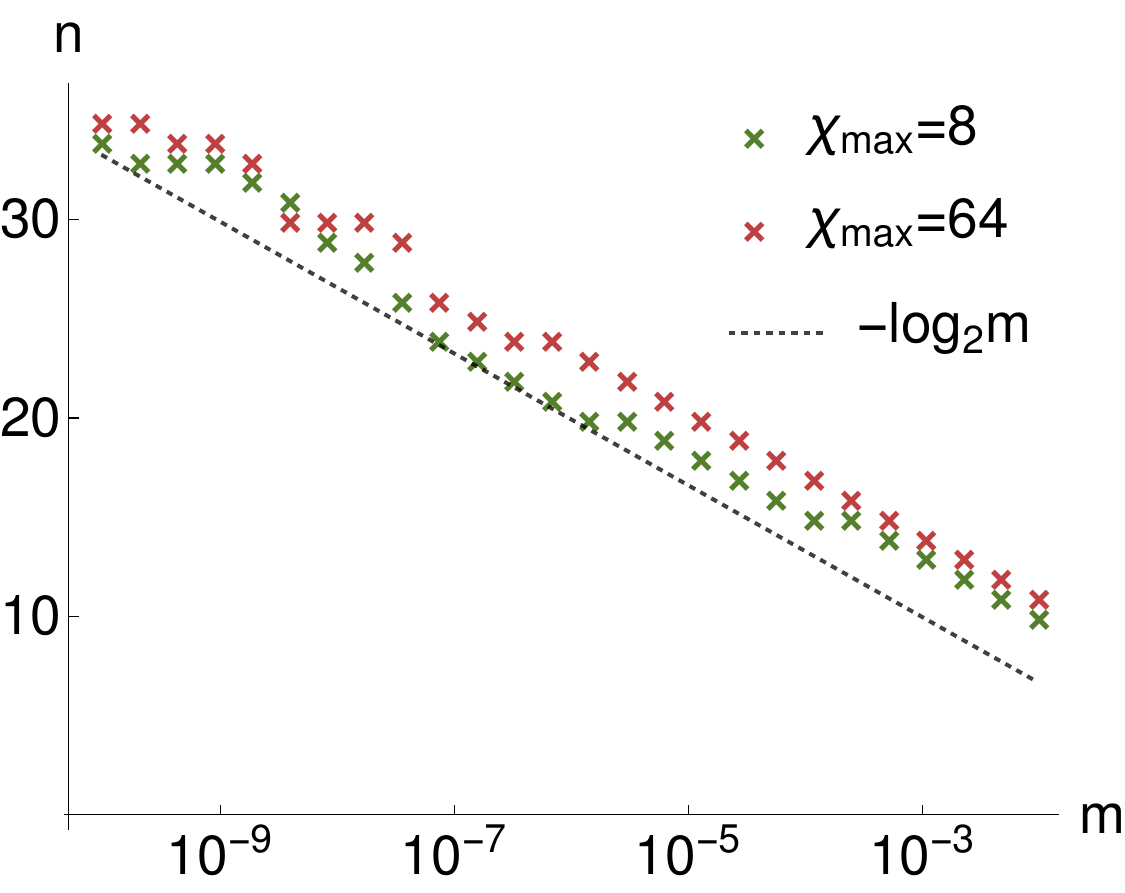}
\end{center}
\vspace{-2mm}
\caption{\label{fig:CDL2}  Left:  Indicator $P_{\rm CDL}$ for $m=10^{-5}$ and $\chi_{\rm max}=24$.
 The step at which it drops coincides with   Fig.\ref{fig:CDL1}-left. Right: RG cycles needed to reach a CDL structure.
The theoretical argument \eqref{cdl} is shown in black.
}
\end{figure}

Let us denote by $\hat B$ the submatrix of $B$ that connects fields on opposite links.
While the pairing of singular values is necessary for CDL, the vanishing of $\hat B$
in two successive gTRG steps is a sufficient condition (see \hyperref[sec:SMC]{SM.C}).
We define
\be
P_{\rm CDL}=
{1 \over \chi} {|| \hat B || \over \lambda_1} \, ,
\label{Pcdl}
\ee
where $||.||$ is the Frobenius norm and $\lambda_1$ the largest singular value of $B$.
The RG evolution of this quantity is plotted in Fig.\ref{fig:CDL2}-left for the example of Fig.\ref{fig:CDL1}-left. It abruptly decreases at the same scale at which the singular values pair up, confirming that the complete CDL structure is realized.

Fig.\ref{fig:CDL2}-right shows the number of RG cycles necessary to attain a CDL IR fixed point using for criterium $P_{\rm CDL}<10^{-7}$.
Similar results are obtained for large and small bond dimensions.
In both cases they are consistent with the scaling argument \eqref{cdl}.
An extrapolation to the massless limit implies $n \rightarrow \infty$ and thus an infinite correlation length. This suggests that the gTRG keeps some
long distance information for any bond dimensions.
The reason behind it could be related with an important feature of the gTRG. It is constructed such that the lattice variables are always fields, which can take arbitrarily large values.
As a consequence the diagonal matrix $S$ in \eqref{svd1}, whose components play the rol of singular values for the gSVD, contains arbitrarily small entries even after truncation.
On the contrary, the ordinary TRG discards the singular values smaller than a chosen cutoff.

{\it Conclusions.-}
We have implemented the  Tensor Renormalization Group method
to compute the partition function of a free boson in two euclidean dimensions.
The guiding  principle is to preserve the gaussian character  of the statistical weights. This led us to modify the  singular value
decomposition to handle continuous degrees of freedom taking unbounded values.   We have obtained very accurate  numerical results  keeping a small  number of fields in the RG iteration procedure.
There is still some residual short range entanglement that give
rise to CDL tensors.  We expect that a version of the TNR along the lines of  references \cite{EV15,EV15b}
would eliminate it completely  reducing  the computational  cost to achieve the same accuracy as it occurs for spin models.
We envisage the generalization of this method to models with interactions.
It would likely require  the use of perturbative techniques.  The final goal is to improve the performance
of the entanglement based RG method  in quantum field theory.

{\it Acknowledgements.}
\begin{acknowledgments}
We would like to thank  M. C. Ba\~nuls,   J. I. Cirac, G. Evenbly, M. Garc\'{\i}a-P\'erez, E. Kim, J.I. Latorre,   C. Pena, S. Ryu, L. Tagliacozzo  and G. Vidal  for conversations.
We acknowledge  financial support from the grants
 FPA2015-65480-P and FIS2015-69167-C2-1-P
(MINECO/FEDER), QUITEMAD+ S2013/ICE-2801 and SEV-2016-0597
of the ``Centro de Excelencia Severo Ochoa'' Programme.
\end{acknowledgments}


\newpage
\onecolumngrid

\section{\large\bf{SUPPLEMENTARY  MATERIAL}}

\appendix

\section{A. gTRG \textsc{algorithm} \label{sec:SMA}}

In order to apply the gTRG algorithm, we first write the bosonic partition function as a contraction of a square tensor network in which each tensor is given by eq.\ref{W}. This tensor $W_1^\phi$ is uniquely identified by a matrix $M_1^\phi$ which encodes all the Boltzmann weights.
\begin{equation}
	W_1^\phi(\phi) = e^{-\frac{1}{2} \phi^T M_1^\phi \phi}
	\ , \qquad
M_1={m \over 2}\,  \id_4 + K \, , \qquad
 K=      \begin{pmatrix}
		\;\;\;2 & -1 & -1 & \;\;\;0 \\
		- 1 &  \;\;\;2 &\;\;\; 0 & -1 \\
		-1 &\;\; \;0 &\;\;2 &\; \;\;1 \\
		\;\;\;0 & -1 & -1 & \; \;\;2
	\end{pmatrix}
	\ ,
\end{equation}
with $\phi= ( \phi_1 , \phi_2 , \phi_3 , \phi_4)$.
Similarly at each step we will have square lattices of tensors $W_n^\varphi$ described by matrices $M^\varphi_n$, where $n$ indicates the RG cycle and $\varphi$ represents the $\phi$ or $\pi$-fields.
The goal of the gTRG algorithm is to compute from $W_n^\varphi$ its coarse-grained version $W_{\tilde n}^{\tilde \varphi}$. If $\varphi = \phi$ then $\tilde \varphi = \pi$ and ${\tilde n}=n$, while if
$\varphi = \pi$  then $\tilde \varphi = \phi$ and  ${\tilde n}= n+1$.  Namely $W_n^\phi \to W_{n}^\pi$ and $W_n^\pi \to W_{n+1}^\phi$.

From now on, when no confusion is possible we just write  $W_n^\varphi = W$ and $W_{\tilde n}^{\tilde \varphi} = \widetilde W$.
Following the TRG, we use the gSVD to split $W(\varphi)=\exp( - {1 \over 2} \varphi^T M \varphi )$ in ``left'' and ``right'' tensors as shown in eq.\eqref{expW}.
Accordingly we separate the fields $\varphi$ in their left and right components $\varphi_L=(\varphi_1,\varphi_2)$ and $\varphi_R=(\varphi_3,\varphi_4)$, where $\varphi_i$ collectively denote all fields that lives in the corresponding lattice link.
$M$ is then decomposed in 4 blocks
\begin{equation}
	M =
		\begin{pmatrix}
				\ \ A & - B \\
				- B & \ \ A
		\end{pmatrix}.
	\label{Mn}
\end{equation}
As we will show, those blocks have further structure and it is possible to decompose them as
\begin{equation}
	A - B =
	\frac{1}{2}
	\begin{pmatrix}
		s & 0 \\
		0 &	s
	\end{pmatrix}
	+
	\begin{pmatrix}
		\ \ a & - a \\
		- a &	\ \ a
	\end{pmatrix}
	\ , \qquad
	B =
	\frac{1}{2} \begin{pmatrix}
		b_{+} + b_{-} \;& \;b_{+} - b_{-} \\
		b_{+} - b_{-} \; &	\;b_{+} + b_{-}
	\end{pmatrix}
	\ ,
	\label{blocks}
\end{equation}
where $a$, $b_{+}$ and $b_{-}$ are $\chi \times \chi$ symmetric and positive semi-definite matrices, and $s$ is a $\chi \times \chi$ diagonal matrix with non-negative entries. The matrices $s$, $a$ and $b_\pm$ act on the fields $\varphi_i$ of each separated lattice link. This structure is verified by the initial weights, where those little blocks are just numbers
\begin{equation}
	s_1 = m^2
	\ , \qquad
	a_1^\phi = b_{+,1}^\phi = b_{-,1}^\phi = 1
	\ .
\end{equation}

The proof proceeds by induction. We assume that the previous structure is realized by $W$. Now we perform the gSVD of $W$ using the SVD of $B$, as explained in the body of the article. Since we have assumed that $b_\pm$ are positive definite, so is $B$, and its SVD reduces to a diagonalization. The diagonalization of $B = U D U^T$ can be computed from the diagonalization of its blocks $b_{\pm} = u_{\pm} d_{\pm} u_{\pm}^T$. The isometries $u_{\pm}$ span the space of non-zero eigenvalues and $d_{\pm}$ is the diagonal matrix with the non-zero eigenvalues of $b_{\pm}$.
The $\tilde \chi \times \tilde \chi$ diagonal matrix $D$ and $2 \chi \times {\tilde \chi}$ isometry $U$ are
\begin{equation}
	D  =
	\begin{pmatrix}
		d_{+} & 0 \\
		0 & d_{-}
	\end{pmatrix}
	\ , \qquad
	U =
	\frac{1}{\sqrt{2}} \begin{pmatrix}
		u_{+} & u_{-} \\
		u_{+} & - u_{-}
	\end{pmatrix}
	\label{iso}
	\ .
\end{equation}
At this point, if the number of new fields $\tilde \chi$ is too big or some of the eigenvalues in $D$ are too small, we can implement the truncation as explained in the main text.

In the original TRG algorithm, each tensor of the lattice is split in two  $W = V \widetilde V^{\dagger}$. The gTRG algorithm proceeds in the same way. Due to the assumed structure of $W$ we have $V = \widetilde V$, so that
\begin{equation}
	W(\varphi_L, \varphi_R) = \int d \tilde \varphi \; V(\varphi_L,\tilde \varphi) V^\dagger(\tilde \varphi,\varphi_R)
	\ ,
\end{equation}
This relation can be written pictorially as
\begin{equation}
	\vcenter{\hbox{\includegraphics{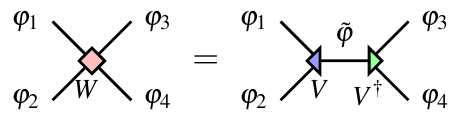}}}
  ,
\end{equation}
where from eqs.\eqref{gaussianSVD}--\eqref{WLR}
\begin{equation}
	V(\varphi_L,\tilde \varphi)
	=
		G(\varphi_L) \, e^{\;i \varphi_L^T U \tilde \varphi} \;  S^{1/2}(\tilde \varphi)
	\ .
\end{equation}

To obtain the new tensor $\widetilde W$ we have to contract a loop of four tensors $V$. Depending on how we label the two halves of each tensor $W$, ``left'' an ``right'', we can have different resulting tensors $\widetilde W$ that are equivalent under a suitable change of fields $\varphi \to -\varphi$. We are going to fix this freedom in such a way that all $\widetilde W$ are equal up to $90^\circ$ rotation, since at the next step they will be split along different axis, and have the structure showed at (\ref{Mn})  and (\ref{blocks}). Our choice can be depicted as
\begin{equation}
	\vcenter{\hbox{\includegraphics{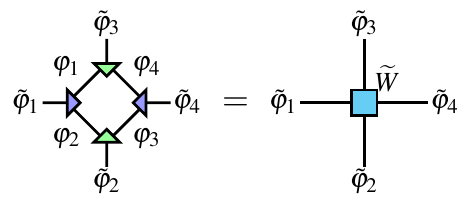}}}
	\label{loop}
\end{equation}
The resulting lattice of tensors preserves the translational and rotational symmetries of the original lattice, but only at the level of plaquettes, as it can be seen in the following figure
\begin{equation}
	\nonumber
	\vcenter{\hbox{\includegraphics{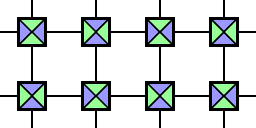}}}
\end{equation}

The new tensor $\widetilde W(\tilde \varphi)$ is given by
\begin{equation}
\widetilde W(\tilde \varphi)=\int \prod_{i=1}^4 d \varphi_i \;
 		V(\varphi_1,\varphi_2;\tilde \varphi_1) \;
 		V^\dagger(\tilde \varphi_2;\varphi_2,\varphi_3) \;
  		V(\varphi_3,\varphi_4;\tilde \varphi_4) \;
  		V^\dagger(\tilde \varphi_3;\varphi_4,\varphi_1) =
		\tilde \rho \; e^{ - \frac{1}{2} \tilde \varphi^T {\widetilde M} \, \tilde \varphi} \ .
\end{equation}
with
\begin{equation}
{\widetilde M} = {1 \over 2}   \id_4 \otimes D^{-1} + C^{\,T} Q^{-1} C \, , \qquad \qquad \tilde \rho =\rho^2 \frac{(2 \pi)^{2 \chi - \tilde \chi}}{\det(D) \det(Q)^{1/2}}  \; .
\end{equation}
The matrix $Q$ collects terms quadratic in $\varphi$ in the exponent of the integrand and $C$ the cross terms in $\varphi$ and $\tilde \varphi$, while $\rho$ is the corresponding factor of $W$.
It is convenient to decompose $C$ in two blocks such that $\varphi^T C \; \tilde \varphi = \varphi^T C_{L} \, \tilde \varphi_L + \varphi^T C_{R} \, \tilde \varphi_R$. We have
\begin{equation}
Q= \id_4 \otimes s+ K \otimes a \ , \qquad
	C_L=
	\begin{pmatrix}
			U &  0 \\
			0 & 0
	\end{pmatrix} - S \begin{pmatrix}
			0 &  0 \\
			0 & U'
	\end{pmatrix}
	\ , \qquad
	C_R =
	\begin{pmatrix}
			0 &  0 \\
			0 & U
	\end{pmatrix} - S \begin{pmatrix}
			U' &  0 \\
			0 & 0
	\end{pmatrix}
	\ , \qquad
	\ .
\end{equation}
where $U'$ is defined as $U$ in \eqref{iso} but substituting $u_-$ by $-u_-$, and the $4\chi \times 4 \chi$ matrix $S$
shifts  $\varphi_i$ to $\varphi_{i-1}$.
Straightforward manipulations show that $\widetilde W$ has the structure described in (\ref{Mn})  and  (\ref{blocks}), with
\begin{equation}
\begin{gathered}
	\tilde s =
		D^{-1}
	\ , \qquad
	\tilde a =
		U^T
		\begin{pmatrix}
				0 & 0 \\
				0 & q_1
		\end{pmatrix}
		U
	\ , \qquad
	\tilde b_{+} =
		U^T
		\begin{pmatrix}
				q_1 & 0 \\
				0 & 0
		\end{pmatrix}
		U
	\ , \qquad
	\tilde b_{-} ={1 \over 2}\,
		U^T
		\begin{pmatrix}
				\,q_0 + q_2 \;\;& q_0 - q_2  \\
				\,q_0 - q_2 \;\; &\;\; q_0+ q_2-2q_1
		\end{pmatrix}
		U
	\ ,
	\label{newblocks}
\end{gathered}
\end{equation}
where $q_j=(s+2 j a)^{-1}$.
The matrix $\tilde s$ is diagonal with non-negative entries. The matrices $\tilde a$, $\tilde b_{+}$ and $\tilde b_{-}$ are symmetric by construction.
They are also positive semi-definite. This is evident for $\tilde a$ and $\tilde b_{+}$ since their eigenvalues are those of $q_1$, which is positive semi-definite because so are $s$ and $a$ by assumption. After some simple algebra, $\tilde b_{-}$ is also shown to be positive semi-definite.

If the eigenvalues of $b_{\pm}$ are all non vanishing, $\tilde \chi=2 \chi$. Since $U$ is then an orthogonal matrix,
the previous expressions make clear that the matrices $\tilde a$ and $\tilde b_{+}$ have half of their eigenvalues equal to zero.
It can be seen that the same result holds for $b_-$. As a result, when we perform a new gTRG iteration the bond dimension does not increase. Moreover $\widetilde U$ is a
$2 {\tilde \chi} \times {\tilde \chi}$ isometry and \eqref{newblocks} does not restrict the number of positive eigenvalues of the new matrices $a$, $b_\pm$. In the generic case, all
of them will be non-vanishing. This property is verified by the initial lattice tensor. Therefore, without truncation, $\chi^\pi_n=2 \chi^\phi_n$ and $\chi^\phi_{n+1}=\chi^\pi_n$.

\begin{center}
	\bf Computation of the partition function.
\end{center}

In this article we compute the partition function of square lattices with $L^2$ sites and periodic boundary conditions, with $L=2^S$. After each gTRG step, the number of sites is reduced by $1/2$. Therefore, after $S-1$ RG steps our lattice only have $4$ sites and there are only two tensors left. Then, performing another gTRG transformation the lattice becomes the tensor trace of just one tensor $W_S^\pi$.
\begin{equation}
	Z=
	\tTr \, W_S^\pi
	=
		\int d \pi_1 d \pi_2 \; W_S(\pi_1,\pi_2,\pi_2,\pi_1)
	=
	\vcenter{\hbox{\includegraphics{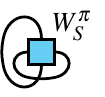}}}
	\ .
\end{equation}
It is important to take into account that the definition of the tensor $W_S^\pi$ in the last step is special, since we are not free to arrange the loop of tensors as in (\ref{loop}). Instead, we are forced to use a disposition in which $V_{S}^\phi$ and $(V_{S}^\phi)^\dagger$ are placed at opposite sides, as in the following figure
\begin{equation}
	\vcenter{\hbox{\includegraphics{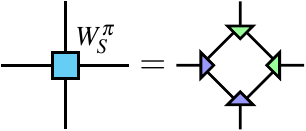}}}
	\ .
\end{equation}

%
%
%

\vspace{1cm}
\section{B. \textsc{Details of the truncation} \label{sec:SMB}}

In order to minimize the numerical error, the gTRG discards singular values of $B$ below a given threshold $\epsilon$.
Without truncation, the singular values of $B^\phi_n$ follow an approximately exponential distribution with smaller values added at each step, see Fig.\ref{fig:svB}-left. If we allow $\chi_{\rm max}$ large enough, at some point some of them will be smaller than $\epsilon$. Using the value $\epsilon = 10^{-11}$, this happens when $\chi_{\rm max} > 22$ for $m < 0.1$,
and at smaller $\chi_{\rm max}$ for bigger masses. Truncations which involves $\epsilon$ have relevant differences with those in which $\epsilon$ plays no role.
In the latter case truncation is only triggered when the maximal bond dimension $\chi_{\rm max}$ is reached. Before that, the bond dimensions doubles in the gTRG steps that lead from $\phi$ to $\pi$-fields and remains constant when transforming from $\pi$ to $\phi$-fields. Therefore
\begin{equation}
	\chi^\pi_{n-1} = \chi^\phi_n = {1 \over 2} \chi^\pi_{n}
	\ .
\end{equation}
On the contrary, a typical sequence of bond dimensions which involves $\epsilon$ is
\begin{equation}
		\{ \chi^\phi_1 ,\, \chi^\pi_1 ,\, \chi^\phi_2 ,\, \chi^\pi_2 ,\, \dots \}
	=
		\{ 1,2,\,2,\,4,\,4,\,8,\,8,\,16,\,16,\,22,\,30,\,35,\,41,\,46,\,54,\,60,\,64,\,64,\,64 ,\, \dots \} \, ,
	\label{chis}
\end{equation}
corresponding to $m = 10^{-6}$ and $\chi_{\rm max} = 64$.  Instead of at once, the maximal bond dimension is now attained in successive steps.

\begin{figure}[h]
	\begin{center}
		\includegraphics[width=6cm]{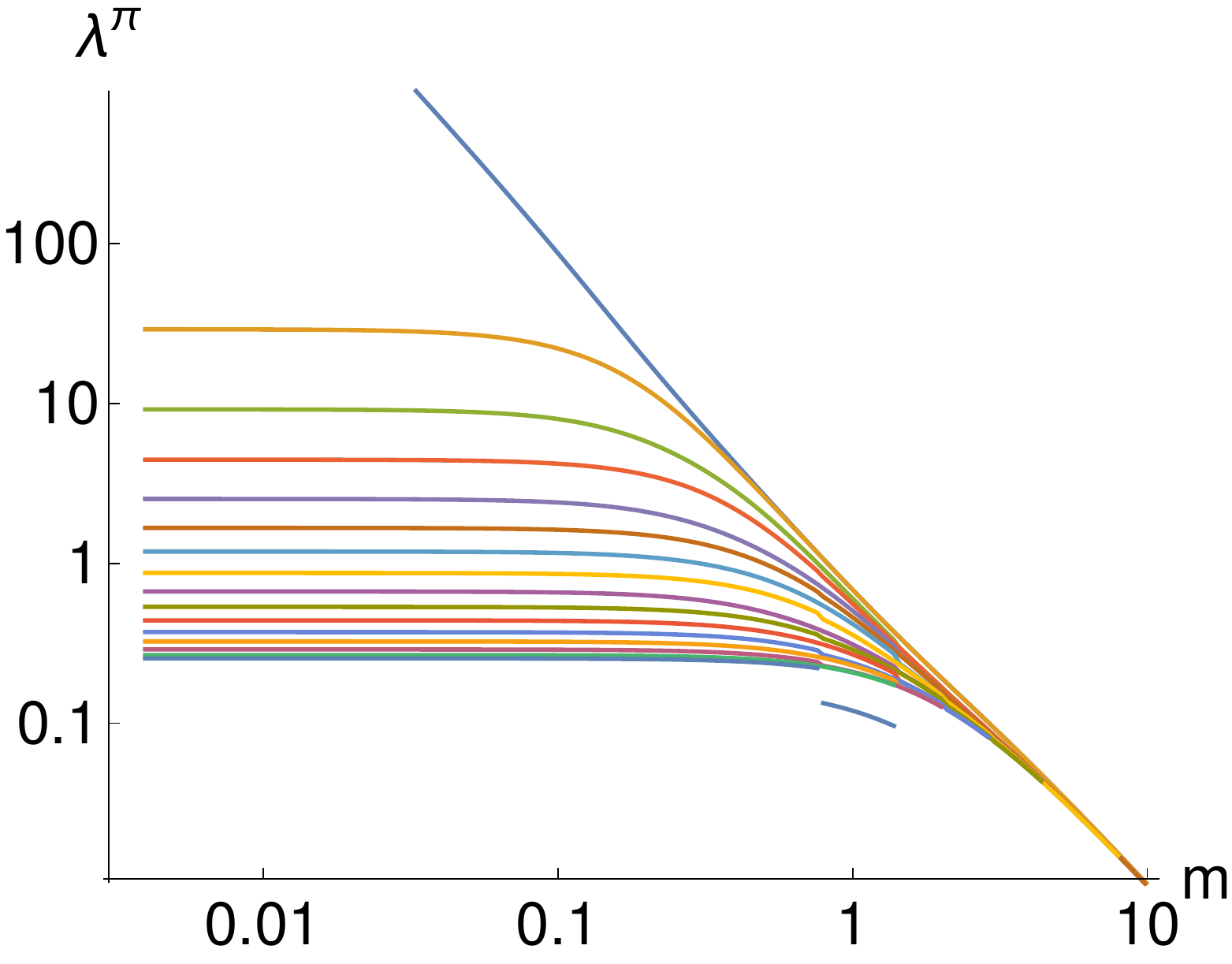}
		\hspace{1cm}
		\includegraphics[width=6cm]{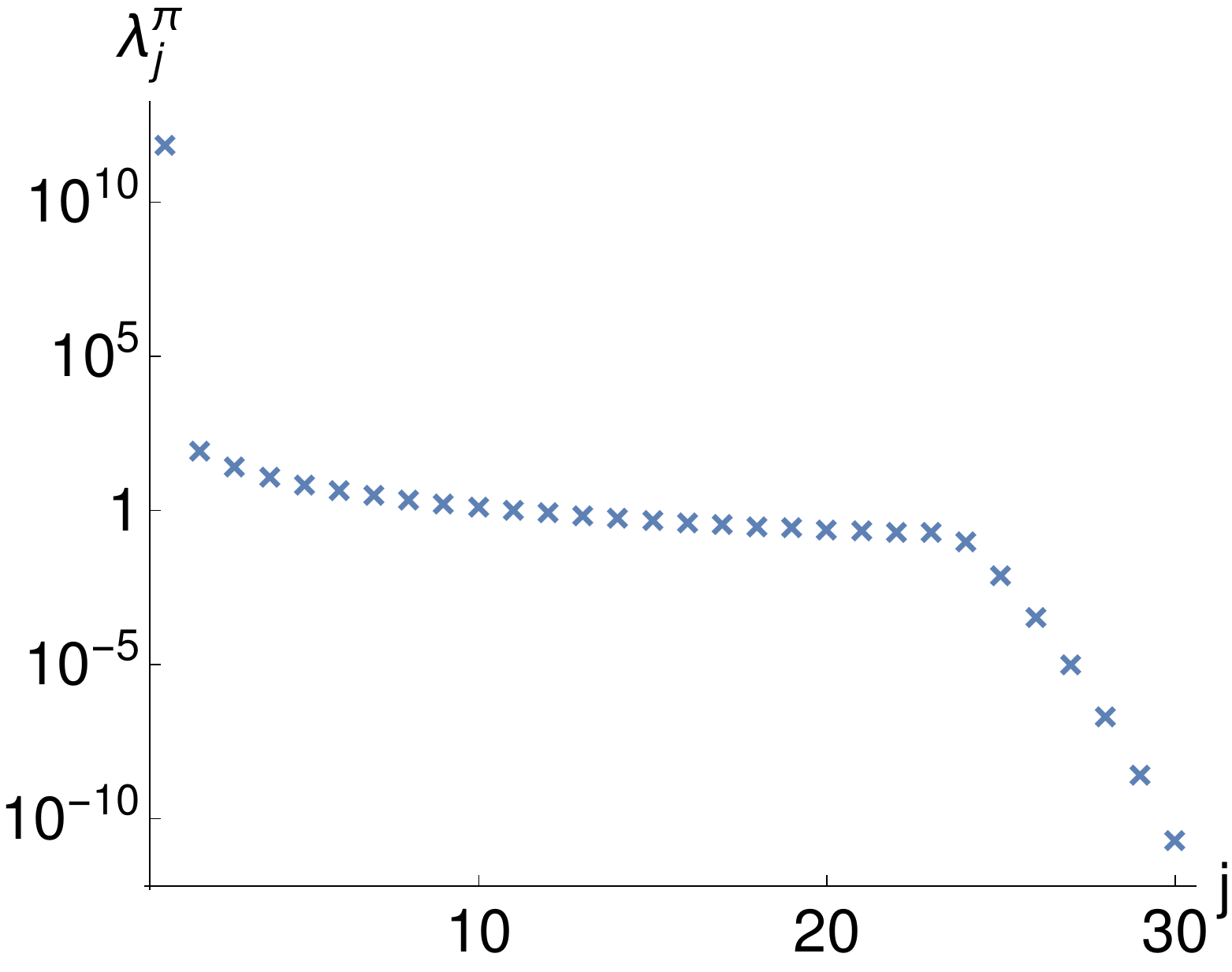}
	\end{center}
	\vspace{-5mm}
	\caption{\label{fig:svpi}
		Left: Singular values of $B^\pi_4$ without truncation. For small masses the largest singular value is approximately $m^{-2}$. For large masses they all converge to $m^{-2}$.
		Right: Singular values of $B^\pi_5$ for $m=10^{-6}$.
	}
\end{figure}

Before truncation, the matrices $B^\pi_n$ have quite different properties from $B_n^\phi$: {\it i)} half of their singular values are zero, {\it ii)} those non-vanishing stay above ${\cal O}(1)$ values.
The singular values of $B^\pi_4$ are shown for illustration in Fig.\ref{fig:svpi}-left. Once $\epsilon$ triggers truncation $\chi_n^\pi < 2 \chi_n^\phi$, as seen in \eqref{chis}.
Moreover, the two previous properties of $B_n^\pi$ are not satisfied anymore. More than half of its singular values are now positive. The largest $\chi^\pi_n$ of them behave as before. The new ones instead decay in an approximately exponential way, similar to those of the $\phi$-lattices. In Fig.\ref{fig:svpi}-right we show the
singular values of the matrix $B_5^\pi$ associated to \eqref{chis}. We observe that the first $\chi^\pi_5=22$ singular values stay above ${\cal O}(1)$, while the next ones strongly decay. A total number of $\chi_6^\phi=30$ survive the $\epsilon$ cutoff. Hence, after truncation is triggered $\chi_n^\phi>\chi_{n-1}^\pi$.

The resulting stepwise pattern of reaching the maximal bond dimension has important consequences in the performance of the gTRG. Fig.\ref{fig:svB}-right shows that it
lowers the numerical cost of improving the precision with respect to cases where $\epsilon$ does not intervene. Interestingly, this turns out to rely on the possibility of having $\chi_n^\phi>\chi_{n-1}^\pi$.
Indeed, we have checked that restricting the bond dimensions to only increase in the $\phi$ to $\pi$ transformations clearly worsens the results.

%
%
%

\vspace{1cm}
\section{C. CDL \textsc{structure} \label{sec:SMC}}

In this section we explain the details of the corner double line (CDL) structure that appears in the gTRG algorithm. The internal structure of the CDL tensors is given by
\begin{equation}
	\nonumber
	e^{-\frac{1}{2} \varphi^T M_{\rm CDL} \varphi}
	\propto
	\vcenter{\hbox{\includegraphics{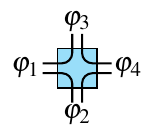}}}
	\ .
\end{equation}
where the internal lines represent cross-terms between the corresponding fields in the exponent. The matrix $M_{\rm CDL}$ factorizes thus in the tensor product
of four equal blocks
\begin{equation}
	M_{\rm CDL} =
		\id_{4} \otimes m_{\rm CDL}
			\ , \qquad m_{\rm CDL}= \id_2 \otimes h + \begin{pmatrix}
			\;\;\; 1 & -1 \\
			-1 & \;\;\; 1
		\end{pmatrix} \otimes k \, , \qquad m_{\rm CDL} \sim
		\vcenter{\hbox{\includegraphics[width=.6cm]{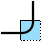}}}
	\label{MCDL}
\end{equation}
where $h$ and $k$ are $\chi/2 \times \chi/2$ symmetric, positive definite real matrices.

In terms of the definitions introduced in \hyperref[sec:SMA]{SM.A}, CDL requires: {\it i)} $b_+=b_-$, {\it ii)} half of the eigenvalues of $a$ and $b_+$ are zero, {\it iii)} the subspaces spanned by the eigenvectors of $a$ and $b_+$ with non-zero eigenvalues are orthogonal, {\it iv)} the mass matrix $s$ does not connect these subspaces.
We will now show that if the submatrices $b_\pm$ coincide in two consecutive gTRG steps, or equivalently, a RG cycle, then the full CDL structure is realized.
The indicator $P_{\rm CDL}$ defined in \eqref{Pcdl}, where $\hat B=(b_+-b_-)/2$, measures the deviation from this condition.
Following the notation of \hyperref[sec:SMA]{SM.A}, we label two consecutive gTRG steps with indices $n$ and $\tilde n$ and their associated fields by $\varphi$ and $\tilde \varphi$. We assume $b_+=b_-$ and $\tilde b_+=\tilde b_-$. The matrix $b \equiv b_+=b_-$ decomposes as $u d u^T$, where the $\tilde \chi/2 \times \tilde \chi/2$ diagonal matrix $d$ collects its positive eigenvalues and $u$ is a $\chi \times \tilde \chi/2$ isometry. Using \eqref{newblocks} and further applying the following change of basis to the fields in each lattice link
\begin{equation}
		\id_{\tilde \chi/2} \otimes \frac{1}{\sqrt{2}}
		\begin{pmatrix}
			1 & \;\;\;1 \\
			1 & -1
		\end{pmatrix}
	\ ,
\end{equation}
we obtain
\begin{equation}
	\tilde s =
	\begin{pmatrix}
		d^{-1} & 0 \\
		0 & d^{-1}
	\end{pmatrix}
	\ , \qquad
	\tilde a =
	\begin{pmatrix}
		0 & 0 \\
		0 & u^T s^{-1} u
	\end{pmatrix}
	\ , \qquad
	\tilde b \equiv \tilde b_+=\tilde b_-=
	\begin{pmatrix}
		u^T s^{-1} u & 0 \\
		0 & 0
	\end{pmatrix}
	\ .
	\label{newcdl}
\end{equation}
These matrices clearly satisfy all the requirements for CDL, and lead to \eqref{MCDL} with $h=d^{-1}$ and $k=u^T s^{-1} u$.

The CDL structure is a fixed point of the gTRG algorithm. Let us perform a gTRG iteration taking as starting point \eqref{newcdl}. The non-zero block of the matrices $a$ and $b$ has maximal rank and thus the new bond dimension is again $\chi_{n+1}=\tilde \chi$. This implies that
$\tilde b= \tilde u \tilde d \tilde u^T$, where $\tilde u^T = \begin{pmatrix} v^T & 0 \end{pmatrix}$ and $v$ is the orthogonal matrix that diagonalizes $u^T s^{-1} u$. The building blocks
of the new tensors, $\tilde q_j=(\tilde s + 2j \tilde a)^{-1}$ defined in \eqref{newblocks}, satisfy
\be
\tilde u^T \tilde q_j \, \tilde u= v^T d\, v  \, .
\ee
Therefore $(b_+)_{n+1}=(b_-)_{n+1}$, and the complete CDL structure is realized with $h=\tilde d^{-1}$ and $k=v^T d v$. A new gTRG iteration leads to $h=d^{-1}$ and $k=v \tilde d v^T$, showing that that a RG cycle leaves invariant the exponent of the gaussian weights. Interestingly, a gTRG step exchanges the roles of $h$ and $k$.

%
%
%

\vspace{1cm}
\section{D. \textsc{Exact results and relation with Conformal Field Theory} \label{sec:SMD}}

Let us consider a lattice $L_1 \times L_2$ and real scalar fields $\phi_{ij}, i=1, \dots, L_1, j=1, \dots, L_2$.
The  partition function   is given by
\beq
Z= \int \prod_{ij} d \phi_{ij} \; e^{- S[\phi]} \; ,
\label{s1}
\eeq
with
\beq
S = \frac{1}{2} \sum_{i=1}^{L_1} \sum_{j=1}^{L_2}
\left[  (\phi_{ij}-\phi_{i + 1 j})^2 \,+\,  (\phi_{ij}-\phi_{i j+1})^2 \,+\, m^2 \phi_{ij}^2 \right] \, .
\label{s2}
\eeq
Let us make the Fourier transform
\beq
\phi_{j_1 j_2} = \frac{1}{ \sqrt{L_1 L_2}} \sum_{k_1, k_2} e^{ i (k_1 j_1 + k_2 j_2)} \hat{\phi}_{k_1 k_2} \, ,
\label{s3}
\eeq
where the periodic boundary conditions imply
\beq
k_i = \frac{ 2 \pi n_i}{L_i} \; (n_i =1, \dots, L_i),  \; i=1,2  \, ,
\label{s4}
\eeq
and the reality condition reads
\beq
\hat{\phi}^*_{k_1 k_2} = \hat{\phi}_{- k_1 - k_2} \, .
\label{s5}
\eeq
In momentum space the action becomes
\beq
 S =  \frac{1}{2} \sum_{k_1, k_2} \left( 4 \sin^2 \frac{k_1}{2} + 4 \sin^2 \frac{k_2}{2}  + m^2 \right) \hat{\phi}_{k_1 k_2}  \hat{\phi}_{k_1 k_2}^*   \, .
 \label{s6}
 \eeq
 Performing  the gaussian  integration yields
 \barray
 Z(L_1, L_2)
& = &  \left( 2 \pi \right)^{L_1 L_2/2}
 \prod_{n_1, n_2}  \left( 4 \sin^2 \frac{\pi n_1}{L_1} + 4 \sin^2 \frac{\pi n_2}{L_2}  + m^2 \right)^{- 1/2}
 \label{s7}
 \earray

\vspace{1cm}
\begin{center}
	\textbf{Relation with CFT}
\end{center}

\begin{figure}[h!]
 \includegraphics[width=8.5cm]{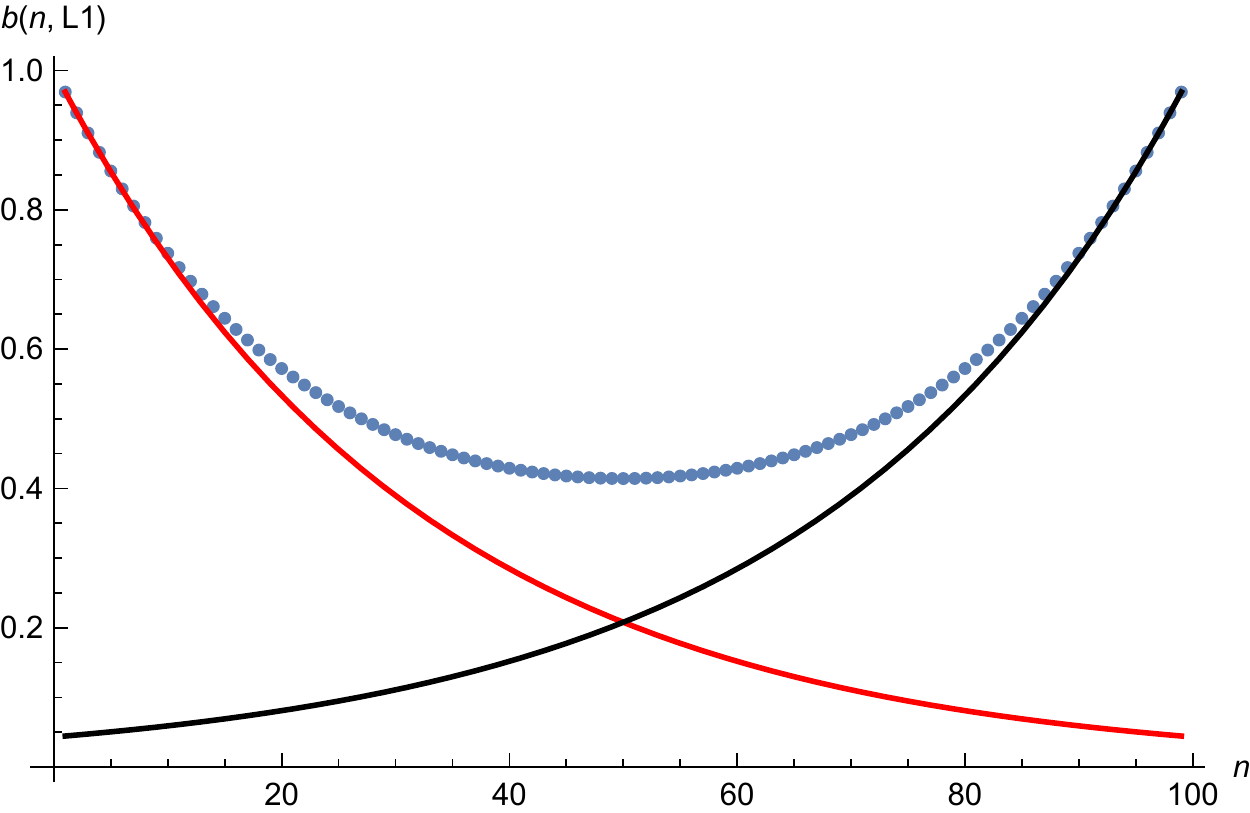}
 \caption{Plot of the function $b(n, L_1)$ for $n=1, \dots, L_1-1$ and  $L_1= 100$.
 The red  curve is $e^{ - \pi n_1/L_1}$ and the blue curve is $e^{ - \pi (L_1 - n_1)/L_1}$. }
 \label{plot1}
\end{figure}

 In the limit $m \rightarrow 0$,  we can approximate   eq.(\ref{s7}) by
 \barray
Z(L_1, L_2)  &  \simeq  &  \frac{2}{m} \left( \frac{ \pi}{ 2 } \right)^{L_1 L_2/2}
 \prod_{(n_1, n_2) \neq (L_1, L_2)}  \left(  \sin^2 \frac{\pi n_1}{L_1} +  \sin^2 \frac{\pi n_2}{L_2} \right)^{- 1/2}  \, .
\label{s8}
 \earray
 We will  compute this product in the limit $L_1, L_2 \gg 1$, keeping the ratio
 $L_2/L_1$ constant. For this purpose we shall employ the following formula
\beq
\prod_{n=1}^L ( x^2 + \sin^2  \frac{ \pi n }{L}  )  = ( 2^{1- L} \sinh( L \,  {\rm arcsinh} \; (x))^2 \, ,
\label{s9}
\eeq
that using
\beq
 {\rm arcsinh} \; (x) = \ln ( x + \sqrt{1+ x^2} ) \, ,
 \label{s10}
 \eeq
becomes
\beq
\prod_{n=1}^L ( x^2 + \sin^2 \frac{ \pi n }{L}  )  = 2^{- 2 L}  \left[ ( x + \sqrt{1+ x^2})^L - ( x + \sqrt{1+ x^2})^{-L} \right]^2    \,  .
\label{s11}
\eeq
Let us write eq.(\ref{s8}) as
 \barray
Z(L_1, L_2)  &  \simeq  &
  \frac{2}{m} \left( \frac{ \pi}{ 2 } \right)^{L_1 L_2/2}
 \prod_{(n_1, n_2) \neq (L_1, L_2)}  a(n_1, n_2)  \, ,
 \label{s12}
 \earray
 where
 \beq
 a(n_1, n_2) =  \left(  \sin^2 \frac{\pi n_1}{L_1} +  \sin^2 \frac{\pi n_2}{L_2} \right)^{-1/2}  \, .
 \label{s13}
 \eeq
We can split  the product  in (\ref{s12}) as
\beq
A \equiv  \prod_{(n_1, n_2) \neq (L_1, L_2)}  a(n_1, n_2)  = \prod_{n_2=1}^{L_2 -1}   a(L_1, n_2)
\times  \prod_{n_1=1}^{L_1 -1}  \prod_{n_2=1}^{L_2 }   a(n_1, n_2)   \, .
\label{s14}
\eeq
The first factor is given by
\beq
\prod_{n_2=1}^{L_2 -1}   a(L_1, n_2)  = \prod_{n_2=1}^{L_2 -1}  \left(  \sin  \frac{\pi n_2}{L_2} \right)^{-1}   =
\left(  2^{ 1 - L_2} L_2 \right)^{-1} \, ,
\label{s15}
\eeq
while the second factor can be obtained  using (\ref{s11}),
\barray
 \prod_{n_1=1}^{L_1 -1}  \prod_{n_2=1}^{L_2 }   a(n_1, n_2) &  =   &
 \prod_{n_1=1}^{L_1 -1}  \prod_{n_2=1}^{L_2 }    \left(  \sin^2 \frac{\pi n_1}{L_1} +  \sin^2 \frac{\pi n_2}{L_2} \right)^{-1/2}
\label{s16} \\
& = &   2^{L_2(L_1-1)}  \prod_{n_1=1}^{L_1 -1}   \left[ ( x_{n_1}  + \sqrt{1+ x_{n_1}^2})^{L_2} - ( x_{n_1} + \sqrt{1+ x_{n_1}^2})^{-L_2} \right]^{-1}
\nonumber  \\
& = & 2^{L_2(L_1-1)}  \prod_{n_1=1}^{L_1 -1}     ( x_{n_1}  + \sqrt{1+ x_{n_1}^2})^{- L_2} \times
\prod_{n_1=1}^{L_1 -1}   \left[ 1 - ( x_{n_1} + \sqrt{1+ x_{n_1}^2})^{-2 L_2} \right]^{-1}  \, ,
\nonumber
\earray
where
\beq
x_{n_1} = \sin \frac{ \pi n_1}{L_1}   \, .
\label{s17}
\eeq
Combining eqs.(\ref{s12}), (\ref{s15}) and (\ref{s16}) yields
 \barray
Z(L_1, L_2)  &  \simeq  &  \frac{ ( 2 \pi)^{ \frac{1}{2} L_1 L_2} }{m L_2}
 \prod_{n_1=1}^{L_1 -1}     ( x_{n_1}  + \sqrt{1+ x_{n_1}^2})^{- L_2} \times
\prod_{n_1=1}^{L_1 -1}   \left[ 1 - ( x_{n_1} + \sqrt{1+ x_{n_1}^2})^{-2 L_2} \right]^{-1}  \, .
 \label{s18}
 \earray
Let us  define
\beq
b(n_1, L_1) = ( x_{n_1} + \sqrt{1+ x_{n_1}^2})^{-1}
=  -  \sin \frac{ \pi n_1}{L_1}   + \sqrt{ 1 + \sin^2  \frac{ \pi n_1}{L_1} }  \, .
\label{s19}
\eeq
Fig. \ref{plot1} shows that for $L_1 \gg 1$, the values of this function
 near 1  can be approximated by
\beq
b(n_1, L_1)  \simeq
\left\{ \begin{array}{ll}
e^{ - \pi n_1/L_1} , & n_1 \ll L_1 \\
e^{ - \pi (L_1 - n_1)/L_1} , & n_1 \simeq  L_1 \\
\end{array}
\right.  \, .
\label{s20}
\eeq
These  analytic expressions can be derived from eq.(\ref{s19}).  Hence in the limit $L_1, L_2 \gg 1$, with $L_2/L_1$ constant,
we find
\beq
\prod_{n_1=1}^{L_1 -1}   \left[ 1 - ( x_{n_1} + \sqrt{1+ x_{n_1}^2})^{-2 L_2} \right]^{-1}
\simeq   \prod_{n_1=1}^{L_1 -1}   \left( 1 - e^{ - 2 \pi L_2 n_1/L_1}  \right)^{-2}   \simeq   \prod_{n=1}^{\infty}   \left( 1 - q^n  \right)^{-2}  \, ,
\label{s21}
\eeq
where
\beq
q= e^{ - 2 \pi L_2/L_1} \, .
\label{s22}
\eeq
The exponent 2 in eq.(\ref{s21}) comes from the terms  around $n_1 \simeq L_1$
that contribute with the same amount as those near $n_1 \ll L_1$.

 Let us now evaluate the first product  in eq.(\ref{s18})
 \barray
 \prod_{n=1}^{L_1 -1}     ( x_{n}  + \sqrt{1+ x_{n}^2})^{- L_2}  =
 {\rm exp} \left( - L_2 \sum_{n=1}^{L_1 -1}  f(n) \right)  \, ,
 \label{s23}
 \earray
where
\beq
f(n) = \ln  ( x_{n}  + \sqrt{1+ x_{n}^2})   = \ln \left( \sin \frac{ \pi n}{L_1}  + \sqrt{1+   \sin^2 \frac{ \pi n}{L_1} } \right)  \, .
\label{s24}
\eeq
To approximate  the sum (\ref{s23}), we use the Euler-MacLaurin formula
\beq
\sum_{n=1}^{L_1-1} f(n) = \int_{0}^{L_1} dn \, f(n) -  \frac{ f(0) +  f(L_1)}{2} + \frac{1}{12} ( f'(L_1) - f'(0))  + \dots
\label{s25}
\eeq
and compute the various terms
\barray
\int_{0}^{L_1} dn \, f(n) & = &   L_1  \int_{0}^1 dx \; \ln \left( \sin (\pi x) + \sqrt{ 1 + \sin^2(\pi x)}  \right) =     \frac{ 2 G}{\pi} L_1 \, ,
\label{s26}
\earray
where $G$ is the Catalan constant.  The rest of the quantities are given in the limit $L_1 \gg 1$ by
\barray
f(0) & = &  f(L_1) =  0 , \label{s27}   \\
f'(0) & = & - f'(L_1) =    \frac{ \pi}{L_1} + O(L_1^{-3})  \nonumber  \, .
\earray
Therefore
\beq
\sum_{n=1}^{L_1-1} f(n) \simeq  \frac{ 2 G}{\pi}  L_1   - \frac{ \pi}{ 6 L_1}  \, ,
\label{s28}
\eeq
which plugged into  eq.(\ref{s23}) yields,
 \barray
 \prod_{n=1}^{L_1 -1}     ( x_{n}  + \sqrt{1+ x_{n}^2})^{- L_2}  =
 {\rm exp} \left[  -   \frac{2 G}{\pi}  L_1 L_2  + \frac{ \pi L_2}{6 L_1}  \right]  =
  {\rm exp} \left[  -  \frac{2 G}{\pi}  L_1 L_2    \right]   q^{ - \frac{1}{12}}  \, .
 \label{s29}
 \earray
Collecting  terms, eq.(\ref{s18}) becomes
 \barray
Z(L_1, L_2)  &  \simeq  &  \frac{ ( 2 \pi)^{ \frac{1}{2} L_1 L_2} }{m L_2}
  {\rm exp} \left[  - \frac{2 G}{\pi}  L_1 L_2   \right]   q^{ - \frac{1}{12}}
  \prod_{n=1}^{\infty}   \left( 1 - q^n  \right)^{-2}   \,  ,
   \label{s30}
 \earray
 that  can be written as
 \beq
 Z(L_1, L_2) \simeq \frac{ e ^{ - L_1 L_2 f_\infty}}{m \sqrt{ L_1 L_2}} \times  Z_{\rm CFT}(\tau)  \, ,
 \label{s31}
 \eeq
 where $f_\infty$ is the free energy per site
 \beq
f_\infty = \frac{ 2 G}{\pi}   - \frac{ \ln (2 \pi)}{2}  \, .
\label{s32}
\eeq
 $Z_{\rm CFT}(\tau)$ is  the partition function of a massless boson on a torus with moduli parameter $\tau$ \cite{CFT}
\beq
Z_{\rm CFT}(\tau) =   \frac{ 1}{ ({\rm Im} \tau)^{1/2} |\eta(q)|^2} ,  \qquad q= e^{ 2 \pi i  \tau} , \quad \tau = i \frac{L_2}{L_1} \, ,
\label{s33}
\eeq
and
\beq
\eta(\tau) = q^{ \frac{1}{24} } \prod_{n=1}^\infty (1 - q^n) \, ,
\label{s33b}
\eeq
is the Dedekind eta function.
Eq.(\ref{s7}) is symmetric under the exchange $L_1 \leftrightarrow L_2$, a condition
that is guaranteed in (\ref{s33}) by the modular invariance  of $Z_{\rm CFT}$
 \barray\;
Z_{\rm CFT}(\tau)   &  =  &  Z_{\rm CFT}(-1/\tau)    \, .
 \label{s34}
 \earray


\vspace{-1mm}

\begin{thebibliography}{99}


\bibitem{V08} F. Verstraete, J.I. Cirac, V. Murg,
``Matrix Product States, Projected Entangled Pair States, and variational renormalization group methods for quantum spin systems'',
Adv. Phys. {\bf 57},143 (2008)


\bibitem{O14} Rom\'an Or\'us, ``A practical introduction to tensor networks:
Matrix product states and projected entangled pair states'',
Ann. Phys. {\bf 349}, 117 (2014).


\bibitem{A88}  I. Affleck, T. Kennedy, E. H. Lieb, and H. Tasaki,
``Valence bond ground states in isotropic quantum antiferromagnets'',
Commun. Math. Phys., {\bf 115}, 477 (1988).

\bibitem{F92}  M. Fannes, B. Nachtergaele, and R. F. Werner,
``Finitely correlated states on quantum spin chains'',
Commun. Math. Phys. {\bf 144}, 443 (1992).

\bibitem{K93}  A. Kl\"{u}mper, A. Schadschneider, and J. Zittartz
``Matrix-product-groundstates for one-dimensional spin-1 quantum antiferromagnets'',
Europhys. Lett. 24, 293 (1993).

\bibitem{O95} S. \"Ostlund and S.  Rommer,
``Thermodynamic Limit of Density Matrix Renormalization'',
Phys. Rev. Lett. {\bf 75}, 3537 (1995).

\bibitem{V03} G. Vidal,
``Efficient Classical Simulation of Slightly Entangled Quantum Computations'',
Phys. Rev. Lett. {\bf 91}, 147902 (2003).

\bibitem{V04}   F. Verstraete, D. Porras, and J. I. Cirac,
``DMRG and periodic boundary conditions: a quantum information perspective'',
Phys. Rev. Lett. {\bf 93}, 227205 (2004).


\bibitem{W92} S. R. White,
``Density matrix formulation for quantum renormalization groups'',
Phys. Rev. Lett. {\bf 69}, 2863 (1992).

\bibitem{D97}  J.Dukelsky, M.A. Martin-Delgado, T. Nishino, G. Sierra,
``Equivalence of the Variational Matrix Product Method and the Density Matrix Renormalization Group applied to Spin Chains'',
Europhys. Lett., {\bf 43}, 457 (1998).


\bibitem{S05}  U. Schollw\"{o}ck,
``The density-matrix renormalization group'',
Rev. Mod. Phys. {\bf 77}, 259 (2005).



\bibitem{VC04}
F. Verstraete and J. I. Cirac,
``Renormalization algorithms for Quantum-Many Body Systems in two and higher dimensions'',
arXiv:cond-mat/0407066v1 (2004).


\bibitem{S98} G. Sierra and M.A. Martin-Delgado
``The Density Matrix Renormalization Group, Quantum Groups and Conformal Field Theory'',
Proceed. Workshop on the Exact Renormalization Group, Faro (Portugal) 1998,
 arXiv:cond-mat/9811170.





\bibitem{V07} G. Vidal,
``Entanglement Renormalization'',
Phys. Rev. Lett. {\bf 99}, 220405 (2007).

\bibitem{G08}  V.  Giovannetti, S.  Montangero, R.  Fazio,
``Quantum MERA Channels'',
Phys. Rev. Lett. {\bf 101}, 180503 (2008).

\bibitem{P09} R. N. C. Pfeifer, G. Evenbly, and G. Vidal,
``Entanglement renormalization, scale invariance, and quantum criticality'',
Phys. Rev. A {\bf 79}, 040301 (2009).








\bibitem{P10}  F. Pollmann, A. M. Turner, E.  Berg, and M.  Oshikawa,
``Entanglement spectrum of a topological phase in one dimension'',
Phys. Rev. B {\bf 81}, 064439 (2010).

\bibitem{C11}  X.  Chen, Z.-C. Gu, and X.-G. Wen,
``Classification of gapped symmetric phases in one-dimensional spin systems'',
Phys. Rev. B {\bf 83}, 035107 (2011).

\bibitem{N11}  N. Schuch, D.  P\'erez-Garc\'{\i}a, and I. Cirac,
``Classifying quantum phases using matrix product states and projected entangled pair states'',
Phys. Rev. B {\bf 84}, 165139 (2011).




\bibitem{Sw12}  B. Swingle,
``Entanglement renormalization and holography'',
Phys. Rev. D {\bf 86}, 065007 (2012).

\bibitem{LS15} J. I. Latorre and G. Sierra,
``Holographic codes'',
arXiv:1502.06618.

\bibitem{PY15}  F.  Pastawski, B.  Yoshida, D.  Harlow, and John Preskill,
``Holographic quantum error-correcting
codes: toy models for the bulk/boundary correspondence'',
J. High Energy Phys. 2015,  {\bf 149},   (2015).

\bibitem{Mo15}  J. Molina-Vilaplana,
``Information geometry of entanglement renormalization for free quantum fields'',
J. High Energy Phys. (2015) 2015:2 (2015).

\bibitem{MN15}  M. Miyaji, T. Numasawa, N. Shiba, T. Takayanagi, K. Watanabe,
``cMERA as Surface/State Correspondence in AdS/CFT'',
Phys. Rev. Lett. {\bf 115}, 171602 (2015)



\bibitem{CK17}  P.  Caputa, N.  Kundu, M.  Miyaji, T.  Takayanagi, and K.  Watanabe,
``Liouville action as path-integral complexity: from continuous tensor networks to AdS/CFT'',
J. High Energy Phys. 2017, {\bf 97}  (2017).




\bibitem{N95} T. Nishino.
``Density Matrix Renormalization Group Method for 2D Classical Models''.
J. Phys. Soc. Jpn., {\bf 64}, 3598 (1995).




\bibitem{M05}  V. Murg, F. Verstraete, and J. I. Cirac.
``Efficient evaluation of partition
functions of frustrated and inhomogeneous spin systems''.
Phys. Rev. Lett., {\bf 95}, 057206  (2005).



\bibitem{LN07}
M. Levin and C. P. Nave,
``Tensor Renormalization Group Approach to Two-Dimensional Classical Lattice Models'',
Phys. Rev. Lett. {\bf 99}, 120601 (2007).

\bibitem{K66}  L.P. Kadanoff, ``Scaling laws for Ising models near Tc'',
Physics (Long Island City, N.Y.) 2, 263 (1966).

\bibitem{W75}  K.G. Wilson, Group and Critical Phenomena. I. Renormalization Group and the Kadanoff Scaling Picture,
Phys. Rev. B {\bf 4}, 3174 (1971).




\bibitem{GW09}
Z.-C. Gu and X.-G. Wen,
``Tensor-entanglement-filtering renormalization approach and symmetry-protected topological order'',
Phys. Rev. B {\bf 80}, 155131 (2009).


\bibitem{EV15}
G.  Evenbly and G.  Vidal,
``Tensor Network Renormalization'',
Phys. Rev. Lett. {\bf 115}, 180405 (2015);

\bibitem{EV15b}
G.  Evenbly and G.  Vidal,
``Tensor network renormalization yields the multi-scale entanglement renormalization ansatz'',
Phys. Rev. Lett. {\bf 115}, 200401 (2015).





\bibitem{V10}  F. Verstraete and J. I. Cirac,
``Continuous matrix product states for quantum fields'',
Phys. Rev. Lett. {\bf 104}, 190405 (2010).


\bibitem{H13}  J.  Haegeman, T. J. Osborne, H. Verschelde and F. Verstraete,
``Entanglement renormalization for quantum fields in real space'',
Phys. Rev. Lett. {\bf 110}, 100402 (2013).



\bibitem{J15}  D. Jennings, C. Brockt, J. Haegeman, T. J.  Osborne and F.  Verstraete,
``Continuum tensor network field states, path integral representations and spatial symmetries'',
New J. Phys. {\bf 17}, 063039 (2015).


\bibitem{TC18}
A. Tilloy, J. I.  Cirac
``Continuous Tensor Network States for Quantum Fields'',
arXiv:1808.00976.


\bibitem{HV18}  Q.  Hu, A. Franco-Rubio, G. Vidal,
``Continuous tensor network renormalization for quantum fields'',
arXiv:1809.05176.

\bibitem{Sh12} Y. Shimizu,
Tensor renormalization group approach  to a lattice boson model,
Mod.  Phys.  Lett.  A {\bf 27},   1250035  (2012).



\bibitem{CFT}
 P. Di Francesco, P. Mathieu, and D. Senechal, Conformal Field Theory (Springer, New York, 2012).

 \bibitem{A86}   I. Affleck,
 ?Universal Term in the Free Energy at a Critical Point and the Conformal Anomaly?,
 Phys. Rev. Lett. {\bf 56}, 746 (1986).

 \bibitem{B86}  H. W. J. Bl\"{o}te, J.  L. Cardy, and M. P. Nightingale,
``Conformal invariance, the central charge, and universal finite-size amplitudes at criticality'',
Phys. Rev. Lett. {\bf 56}, 742 (1986).










\end{thebibliography}
\end{document}